\documentclass{aa}  

\usepackage{graphicx}
\usepackage{txfonts}
\usepackage{longtable}
\usepackage{array}
\usepackage{multirow}
\usepackage{amsmath}
\usepackage{hyperref}

\begin{document}

   \title{The extremely low-luminosity Type Iax SNe 2022ywf and 2023zgx}

\titlerunning{Extremely faint type Iax SNe 2022ywf and 2023zgx}

   \author{B. Barna\inst{1,2} \and
          D. Bánhidi\inst{1} \and
          T. Szalai\inst{1,3} \and
          J.~P. Anderson\inst{4} \and
          T. Boland\inst{5,6} \and
          K. A. Bostroem\inst{7,8} \and
          T.-W. Chen\inst{9} \and
          J. Farah\inst{10,11} \and
          M. Gromadzki\inst{12} \and
          G. Hosseinzadeh\inst{13} \and 
          D. A. Howell\inst{10,11} \and
          C. Inserra\inst{14} \and
          S. W. Jha\inst{5} \and
          L. Kwok\inst{15} \and
          C. Macrie\inst{5} \and
          C. McCully\inst{10,11} \and
          E. Mochnács\inst{1} \and
          T. E. Müller-Bravo\inst{16,17} \and
          M. Newsome\inst{18} \and
          E. Padilla Gonzalez\inst{10,11} \and
          J. Pearson\inst{7} \and
          T. Petrushevska\inst{19} \and
          D. J. Sand\inst{7} \and
          M. Shrestha\inst{20,21} \and
          N. Smith\inst{7} \and
          S. Srivastav\inst{22} \and
          G. Terreran\inst{10,11} \and
          J. Vinkó\inst{1,23} 
          }

   \institute{Department of Experimental Physics, University of Szeged,
              D\'om t\'er 9, H-6720 Szeged, Hungary
        \and    
            HUN-REN-SZTE Stellar Astrophysics Research Group, 6500 Baja, Szegedi út, Kt. 766, Hungary
        \and
            MTA-ELTE Lend\"ulet "Momentum" Milky Way Research Group, Szent Imre H. st. 112, 9700 Szombathely, Hungary
        \and
            European Southern Observatory, Alonso de Córdova 3107, Vitacura, Casilla 19001, Santiago, Chile
        \and
            Department of Physics and Astronomy, Rutgers, the State University of New Jersey, 136 Frelinghuysen Road, Piscataway, NJ 08854, USA
        \and
            Department of Physics and Astronomy, Purdue University, 525 Northwestern Ave, West Lafayette, IN 47907, USA
        \and
            Steward Observatory, University of Arizona, 933 North Cherry Avenue, Tucson, AZ 85721-0065, USA
        \and
            LSST-DA Catalyst Fellow
        \and
            Graduate Institute of Astronomy, National Central University, 300 Jhongda Road, 32001 Jhongli, Taiwan
        \and
            Las Cumbres Observatory, 6740 Cortona Drive, Suite 102, Goleta, CA 93117-5575, USA
        \and
            Department of Physics, University of California, Santa Barbara, CA 93106-9530, USA
        \and
            Astronomical Observatory, University of Warsaw, Al. Ujazdowskie 4, 00-478 Warszawa, Poland  
        \and
            Department of Astronomy \& Astrophysics, University of California, San Diego, 9500 Gilman Drive, MC 0424, La Jolla, CA 92093-0424, USA
        \and
            Cardiff Hub for Astrophysics Research and Technology, School of Physics \& Astronomy, Cardiff University, Queens Buildings, The Parade, Cardiff, CF24 3AA, UK
        \and
            Center for Interdisciplinary Exploration and Research in Astrophysics (CIERA), 1800 Sherman Ave., Evanston, IL 60201, USA
        \and
            School of Physics, Trinity College Dublin, The University of Dublin, Dublin 2, Ireland
        \and
            Instituto de Ciencias Exactas y Naturales (ICEN), Universidad Arturo Prat, Chile
        \and
            Department of Astronomy, The University of Texas at Austin, 2515 Speedway, Stop C1400, Austin, TX 78712, USA
        \and
            Center for Astrophysics and Cosmology, University of Nova Gorica, Vipavska 11c, 5270 Ajdov\v{s}\v{c}ina, Slovenia
        \and
            School of Physics and Astronomy, Monash University, Clayton, Australia
        \and
            OzGrav: The ARC Center of Excellence for Gravitational Wave Discovery, Australia
        \and
            Astrophysics sub-Department, Department of Physics, University of Oxford, Keble Road, Oxford, OX1 3RH, UK
        \and
            Konkoly Observatory, Research Centre for Astronomy and Earth Sciences, Konkoly Thege Mikl\'os út 15-17, H-1121 Budapest, Hungary
             }


 
  \abstract
   {We present the optical follow-up of SNe 2022ywf and 2023zgx, two examples from the Iax subclass of thermonuclear supernova (SN) events. With peak absolute magnitudes of $M_\mathrm{V} = -13.7$ and $-14.4$ mag, respectively, both objects belong to the extremely low-luminosity (EL) population of the class.}
   {A common origin of SNe in the Iax subclass is still under debate since the distribution of certain observables may indicate that the extremely low-luminosity explosions form a distinct population. We aim to estimate the physical properties of the two EL objects, including mapping the ejecta structure. The results are compared to the predictions of the pure deflagration model with similar luminosity, as well as the common features of other SNe Iax. }
   {We perform spectral tomography on the spectral series of SNe 2022ywf and 2023zgx around their maxima to map the physical properties of the ejecta. Together with the analysis of $BgVriz$ photometry, a wide range of observables can be studied to investigate their distribution against luminosity. The constrained chemical abundances of the ejecta are compared to the predictions of the hydrodynamic simulations with similar peak luminosities. }
   {Constant abundances provide a good match for the distribution of chemical elements for both SNe 2022ywf and 2023zgx. The discrepancies compared to the least luminous pure deflagration model \textit{N5def\_hybrid} are minor, especially at post-maximum epochs. The two SNe also share similar characteristics in their constrained density structures, as well as the evolution of the photosphere.}
  {The analysis supports the assumption that pure deflagration models can reproduce the main characteristics of SNe Iax, even for the low-luminosity population. The presented indirect observational evidence indicates that these objects show similar intrinsic properties to the well-studied, relatively luminous Iax sample and fit into the velocity distribution of the subclass.}
   \maketitle
%

\section{Introduction}

SNe Ia originate from the thermonuclear explosion of C/O white dwarfs (WDs). Although significant exploration has still not confirmed the most probable progenitor and explosion scenario, the remarkable homogeneity of the so-called `normal' SNe makes the Ia class the most important objects for distance estimations \citet{Phillips93}. However, the several subclasses of thermonuclear SNe \citep{Taubenberger17} indicate that there are multiple channels for degenerate WDs; thus, studying these peculiar explosions provides essential information on SN physics.

The most numerous subclass of thermonuclear SNe is formed by the Iax (or, after their prototype object, `2002cx-like') SNe. The volume-limited samples of nearby surveys investigated by \cite{Foley13} indicate a $31_{-13}^{+17} \%$ relative rate of SNe Iax to normal Ia explosions. However, the sample from the recent Data Release 2 of the Zwicky Transient Facility (which included $\sim$3600 and $\sim$1580 thermonuclear SNe in the total and volume-limited samples, respectively) indicates that the relative rate of SNe Iax reached only $\sim$ 4.5\% \citet{Dimitriadis24}. A possible reason for the discrepancy is that the relative rates of sub-luminous SNe Ia can be estimated only by assuming correction factors for the lower probability of detection and classification. This can be critical for SNe Iax with the lowest peak luminosity and shorter light curve widths, adding high uncertainty for population estimates.

The lower peak luminosity of SNe Iax also strongly limits the sample suitable for detailed analysis, in contrast to the normal SNe Ia. Since the first classification \citep[SN 2002cx,][]{Li03}, approximately 110 transients have been identified as type Iax SN according to WISeREP\footnote{WISeREP can be accessed at \url{https://www.wiserep.org/}}, but there are fewer than 20 of them with published well-sampled photometric and spectroscopic datasets. The extreme diversity of the subclass hinders the investigation of the nature of type Iax SNe. Their peak absolute magnitude covers a wide range, from the relatively luminous objects with a peak absolute magnitude of $M_R= -18.60$ mag \citep[SN 2011ay,][]{Szalai15} to extremely faint ones with $M_r=-12.66$ mag \citep[SN 2019gsc,][]{Karambelkar21} in the R/r-band. Similarly, other physical characteristics of these SNe also show extremities, such as the decline rate, spectral velocities, or implied ejecta masses.

Multiple mechanisms have been assumed to explain the peculiar nature of Iax explosions. So far, the pure (or, as it is often called, ``failed'') deflagration scenario has been considered the most promising concept \citep[for a full review, see][]{Lach22}. Hydrodynamic simulations \citep{Jordan12,Kromer13,Long14,Fink14,Kromer15,Leung20,Lach22} have shown that the diverse physical properties of the sample can be reproduced by scaling the strength of the deflagration of a C/O WD. A common feature in all pure deflagration models is that the kinetic energy is not sufficient to fully unbind the WD; thus, the explosion leaves behind a bound remnant. As a further consequence, the ejecta of a SN Iax is expected to have a lower mass and a lower average density than that of normal SNe Ia. However, this scenario cannot account for the least luminous SNe Iax, represented by the prototype SN 2008ha. For this object, a C/O/Ne WD was proposed as a potential progenitor, which successfully decreased the produced energy - and, thus, also the luminosity - of the models \citep{Kromer15}. Note that no other pure deflagration model has produced SNe fainter than $M_\mathrm{r} > -15$ mag, while \cite{Kromer15} presented the hydrodynamic simulation of only one scenario. Since SN 2008ha, even less luminous SNe have been discovered \citep[e.g. SNe 2019gsc, 2021fcg, 2024vjm][]{Tomasella20,Srivastav20,Karambelkar21,Kwok25}. For these objects, the double degenerate merger model presented by \cite{Kashyap18} can reproduce their peak absolute magnitudes and low ejected masses. This scenario results in a failed detonation, providing numerous testable discrepancies compared to the concept of failed deflagration, such as the strong stratification of chemical elements or the relatively higher expansion velocities.

A major breakthrough in understanding SNe~Iax came with the discovery of the progenitor system of SN~2012Z \citep{McCully14,Stritzinger15} in pre-explosion HST images of its host NGC 1309. The luminous blue progenitor makes it highly probable that the system included a helium-star donor and an accreting near-Chandrasekhar mass WD. If this interpretation is correct, it directly confirms that at least some WD SNe come from the single degenerate scenario \citep{McCully14}. In contrast, no luminous companion star has ever been directly observed for a normal SN~Ia, or any other SNe Iax. 


Therefore, the common origin of the extremities of SNe Iax is still under debate, and most studies divide the subclass into luminosity samples. \cite{Srivastav22a} draw the line at peak absolute magnitude $M_\mathrm{V} = - 16$ mag to distinguish the ``faint'' SNe Iax from the rest of the category. \cite{Singh23} argued for the existence of clustering among SNe Iax, as the ``faint'' ($M_\mathrm{r} > - 14.6$ mag) and ``bright'' ($M_\mathrm{r} < - 17.1$ mag) samples showed different correlations between peak luminosity and the $\Delta m_\mathrm{15}$ parameter describing the fading after the peak. 
SNe Iax with an intermediate luminosity range between the two groups presented by \cite{Singh23} have long been known; however, their low numbers and the lack of observational series have made their characterization difficult. The first example with a detailed analysis, SN 2019muj \citep{Barna21}, showed a transitional nature that bridged the luminosity gap between the faint and bright groups. The constrained intrinsic characteristics (e.g., ejecta mass, expansion velocities) and the spectroscopic evolution imply that most of the physical properties and observables change continuously through the entire Iax subclass, instead of clustering. Recent studies of objects with similar luminosities \citep[SNe 2022xlp and 2024pxl,][respectively]{Banhidi25,Singh25} also supported this result. 
Whether or not there are differences in their origin scenarios, labeling various groups of the diverse Iax subclass is useful for referencing the objects with different properties. We suggest, and hereafter use, a three-level (sub-)classification:

\begin{itemize}
    \item Relatively luminous (RL) group: ($M_\mathrm{r} < - 17$ mag), example: SN 2011ay
    \item Intermediate luminous (IL) group: (-14.5 mag > $M_\mathrm{r} > - 17$ mag), example: SN 2019muj
    \item Extremely-low luminosity (EL) group: ($M_\mathrm{r} > - 14.5$ mag), example: SN 2008ha
\end{itemize}

As discussed above, the diverse nature of SNe Iax can be fully understood only by sampling the entire subclass. The cases of SNe 2022ywf and 2023zgx, both peaking around $M_\mathrm{V} \sim -14$ mag, are presented in this study to provide further characteristics in observables and physical properties, particularly of the lowest-luminosity events of this class. The paper is organized as follows. In Sec. \ref{sec:observations} we describe the discovery of the two objects and the properties of their host galaxies, followed by the sources and the properties of the observed datasets. We introduce the techniques used to analyze the photometric and spectroscopic data in Secs. \ref{sec:phot_analysis} and \ref{sec:spec_analysis}, respectively, and describe the inferred results. The conclusions are summarized in Sec. \ref{sec:conclusions}.

\section{Observations}
\label{sec:observations}
\subsection{SN 2022ywf}
\label{sec:22ywf}

SN 2022ywf (RA 01:22:12.20, DEC +00:57:23.55) was discovered by the Zwicky Transient Facility (ZTF) at 59880.18 MJD while still on the rise at $m_\mathrm{r}=19.95$ mag with the last non-detection from 59879.33 MJD with a magnitude limit of 20.4 mag in g-band. The field of SN 2022ywf was also observed by the Asteroid Terrestrial-impact Last Alert System program \citep[ATLAS;][]{Tonry18}, providing a deep, 20.7 mag non-detection limit in cyan-band on 59879.19 MJD. Classification as type Ia SN was based on the first spectrum obtained at 59881.3 MJD \citep{Bostroem22}, which was then refined to be a type Iax with subsequent data \citep{Srivastav22b}. 
The host of SN 2022ywf is NGC 493, an SBc type galaxy. The distance of NGC 493 has been constrained multiple times by following the Tully-Fisher relation. These results converged to the range of 22.0-24.5 Mpc in the past decade. Hereafter, we use the latest $d=22.8 \pm 4.7$ Mpc distance estimation of \citep{Tully16}. The SN appeared at 50" ($\sim5600$ pc) north-east from the center of the host galaxy. Due to its relatively outskirt location in the galactic disk, and because of the non-detection of the Na I D lines at the redshift of NGC 493, we can safely assume negligible host-galaxy reddening.

The ATLAS photometry at the position of SN 2022ywf was obtained through the public forced photometry server\footnote{https://fallingstar-data.com/forcedphot/} \citep{Shingles21}. In addition to the measured fluxes (in microjansky or $\mu$ Jy) and associated errors, the forced photometry server also returns quality metrics including a reduced chi squared for the point spread function (PSF) fitting (labeled chi/N). Individual measurements with an error exceeding $150\,\mathrm{\mu Jy}$ were rejected, along with measurements with chi/N $>3$. ATLAS typically obtains $4 \times 30$\,s exposures spaced within $\sim 1\,$hr. Each quad of individual measurements that passed the quality cuts defined above were combined into a single measurement using a weighted stacking recipe following \cite{Srivastav23}. Stacked measurements with $3\sigma$ significance were considered as detections, and an AB magnitude was derived using $m_{\mathrm{AB}} = -2.5 \times \mathrm{log} (F_{\mathrm{\mu Jy}}^{\mathrm stack}) + 23.9$. The non-detections were converted into $3\sigma$ upper limits using $m_{\mathrm{AB}} > -2.5 \times \mathrm{log} (3 \times \Delta F_{\mathrm{\mu Jy}}^{\mathrm stack}) + 23.9$.

\begin{figure}
\centering 
	\includegraphics[width=1\linewidth]{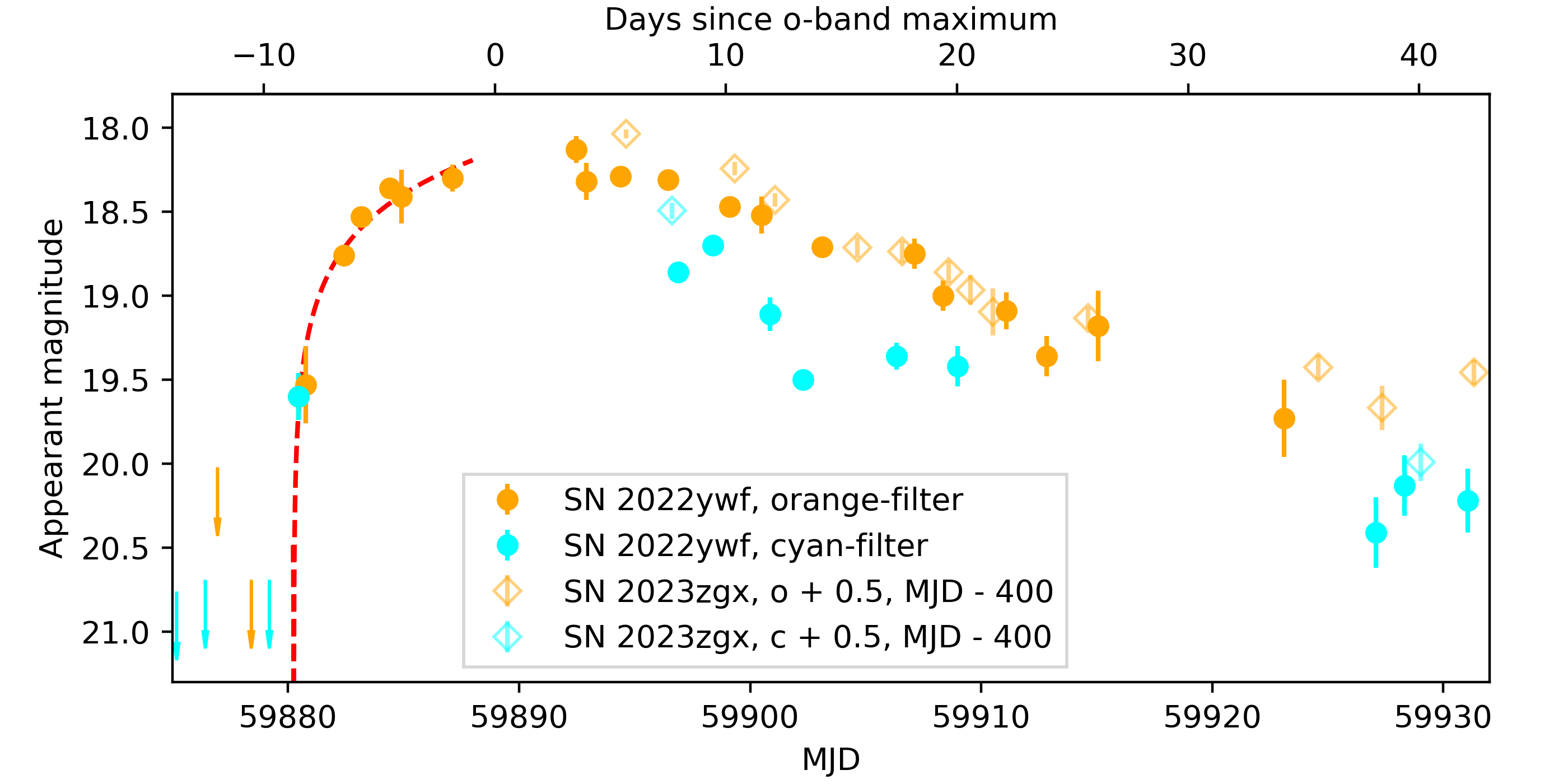}
    \caption{ATLAS photometry of SNe 2022ywf and 2023zgx. The observation dates of SN 2023zgx are shifted with 400 days earlier. The red dashed line show the fit of Eq. \ref{eq:fireball} on the pre-maximum $o$-band data of SN 2022ywf.}
    \label{fig:sn22ywf_atlas}
\end{figure}

\begin{figure}
\centering 
	\includegraphics[width=1\linewidth]{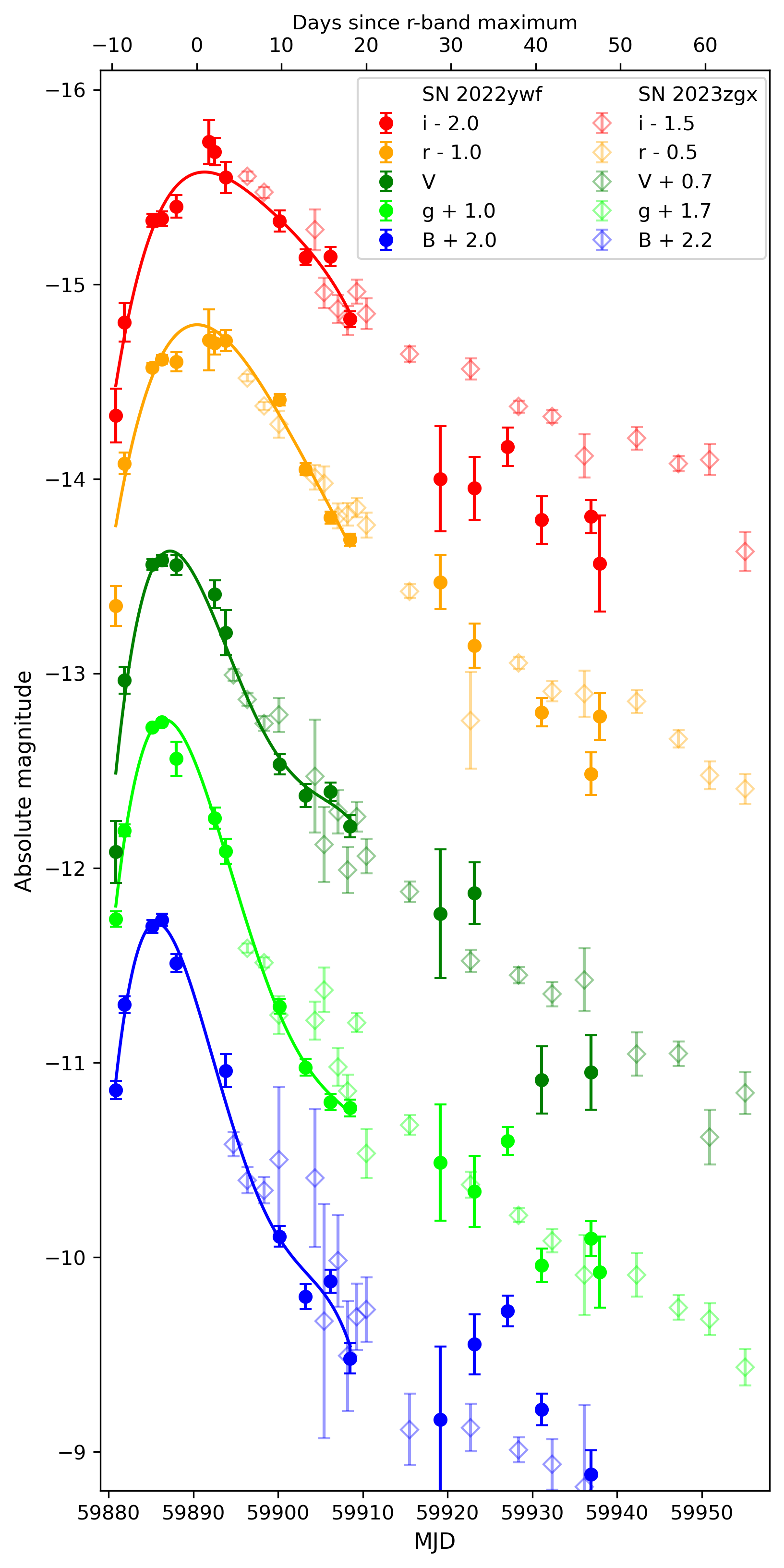}
    \caption{Ground-based photometry of SNe 2022ywf (filled circles) and 2023zgx (fainter diamonds). The horizontal axis represents the observational dates of SN 2022ywf in MJD. For SN 2023zgx, the data points are shifted with 400 days earlier for better comparison. The LC fits with fourth-order polynomial functions are also shown for SN 2022ywf (solid lines). }
    \label{fig:sn22ywf:photometry}
\end{figure}

The Las Cumbres Observatory (LCO) initiated follow-up at 59880.8 MJD, providing $BgVri$ light curves obtained with multiple 1-m telescopes \citep{Brown13} located at Sutherland (South Africa), CTIO (Chile), Siding Spring (Australia), and McDonald (USA) observatories, through the Global Supernova Project (GSP). After +30 days, $BV$ light curves showed highly scattered and uncertain magnitudes, but the $gri$ data are included until 59984 MJD, when the photometric monitoring ceased (Tab. \ref{tab:22ywf_phot}).  The ATLAS survey followed SN 2022ywf in $c$- and $o$-bands. The first detection was at 59880.45 MJD ($m_\mathrm{c}=19.60 \pm 0.14$ mag) and it was regularly observed until 59931.6 MJD, when the target faded below the detection limit (Tab. \ref{tab:ATLAS_22ywf_phot}). The light curves of SN 2022ywf obtained in ATLAS o- and c-bands are shown in Fig. \ref{fig:sn22ywf_atlas}, while the $BgVri$ photometry of SN 2022ywf is compared to that of SN 2023zgx in Fig. \ref{fig:sn22ywf:photometry}.

Optical spectroscopy of SN~2022ywf was obtained by the 9.2-m Southern African Large Telescope (SALT) with the Robert Stobie Spectrograph (RSS) through the Rutgers University program. LCO optical spectra were taken with the FLOYDS spectrographs mounted on the 2-m Faulkes Telescope North and South at Haleakala (USA) and Siding Spring (Australia) through the GSP. Three spectral epochs were observed with the Low Resolution Spectrograph 2 (LRS2) \citep{chonis_lrs2:_2014} mounted on the Hobby-Eberly Telescope \citep[HET,][]{1998SPIE.3352...34R} located at McDonald Observatory. 
Two spectra were obtained with the Southern Astrophysical Research (SOAR) Telescope using the red camera of the Goodman High-Throughput Spectrograph \citep[GHTS;][]{Clemens04} and were reduced using the Goodman Pipeline.\footnote{https://soardocs.readthedocs.io/projects/goodman-pipeline/}.
Additional spectra were observed with the KOSMOS medium resolution spectrograph equipped on the Apache Point Observatory (APO) ARC 3.5m Telescope \citep{}; with the Boller and Chivens Spectrograph (B\&C) on the University of Arizona's Bok 2.3m telescope located at Kitt Peak Observatory; the ESO Faint Object Spectrograph and Camera version 2 \citep[EFOSC2;][]{Buzzoni84} at the 3.6-m New Technology Telescope (NTT) in the frame of the ePESSTO+ collaboration \citep{Smartt15}; and using the SPRAT low resolution spectrograph on the Liverpool Telescope (LT). All optical spectra of SN 2022ywf are shown in Fig. \ref{fig:sn22ywf_spectroscopy}.

\subsection{SN 2023zgx}
\label{sec:23zgx}

SN 2023zgx (RA 12:44:05.00, DEC -05:40:53.47) was discovered by Koichi Itagaki at 60287.8 MJD, likely a few days before its peak, at a visual brightness of 16.9 mag. The classification spectrum, based on which the transient was characterized as a faint type Iax SN, was obtained only four days later \citep{Crimes23}, and photometric follow-up started during its decline. SN 2023zgx was associated with the DDO 142 type S  galaxy. The only distance estimate apart from the Hubble-flow gives $25.4 \pm 5$ Mpc for the distance \citep{Tully88}, which is used in this study. This distance is higher than that calculated from the redshift of 0.0047 ($d_z = 20.2$ Mpc), but still consistent with it. SN 2023zgx was also monitored by the LCO sites, providing $BgVri$ light curves between 60294.7 and 60378.2 MJD. 

The first spectrum of SN 2023zgx was observed with the Gemini Multi-Object Spectrograph mounted at Gemini North \citep[GMOS-N;][]{Hook04} two days before the start of the photometric monitoring. Spectroscopic follow-up was initiated by \href{http://www.pessto.org/}{ePESSTO+} collaboration \citep{Smartt15} with the EFOSC2/NTT and GSP with multiple LCO/FLOYDS spectrographs. Late-time ($t \ +80$ days) spectra were obtained with SALT/RSS between +80 and +90 days after the r-band maximum.

\begin{figure}
\centering 
	\includegraphics[width=1\linewidth]{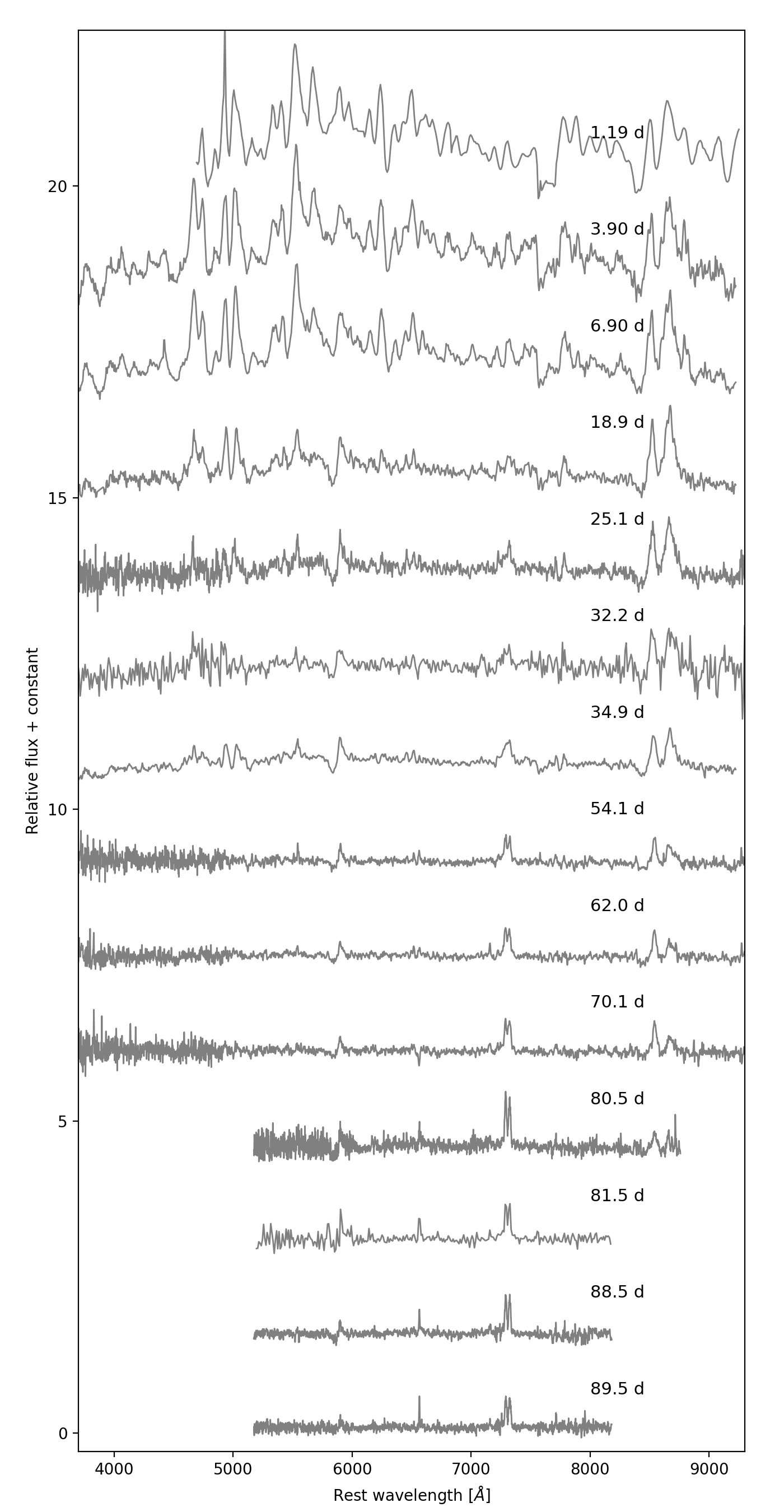}
    \caption{Optical spectra of SN 2023zgx. The epochs
show the days with respect to r-maximum. The observation log of spectra
can be found in Tab. \ref{tab:23zgx_spectra}.}
    \label{fig:sn23zgx_spectroscopy}
\end{figure}

The observed $BgVriz$ photometry of SN 2023zgx can be found in Tab. \ref{tab:23zgx_phot}, while the reduced spectra, corrected for host-galaxy redshift, are shown in Fig. \ref{fig:sn23zgx_spectroscopy}. For further analysis, all spectra were scaled to the photometry using a polynomial function and dereddened for Milky Way extinction.

\section{Photometric analysis}
\label{sec:phot_analysis}



For SN 2022ywf, the photometric coverage is suitable for fitting a low-order polynomial function to the LCs of each filter around the maximum, extending to +20 days. The inferred parameters, such as peak magnitude and $\Delta\,m_\mathrm{15}$ decline rate, are listed in Tab. \ref{tab:phot_fits}. Considering the $\mu=31.79$ mag distance modulus and the MW reddening (assuming no host galaxy reddening based on the lack of Na I D lines) the peak absolute magnitude of SN 2022ywf is $M\mathrm{r} = -13.81 \pm 0.44$ mag, placing the object in the EL group of the Iax subclass. Based on their peak luminosities, the closest relatives to SN 2022ywf are SNe 2008ha and 2019gsc. The polynomial fits provided $T_\mathrm{max} = 59890.4$ MJD) for the moment of maximum light in $r$-band, which is used as a further reference date, similarly to our previous works.

\begin{figure}
\centering 
	\includegraphics[width=1\linewidth]{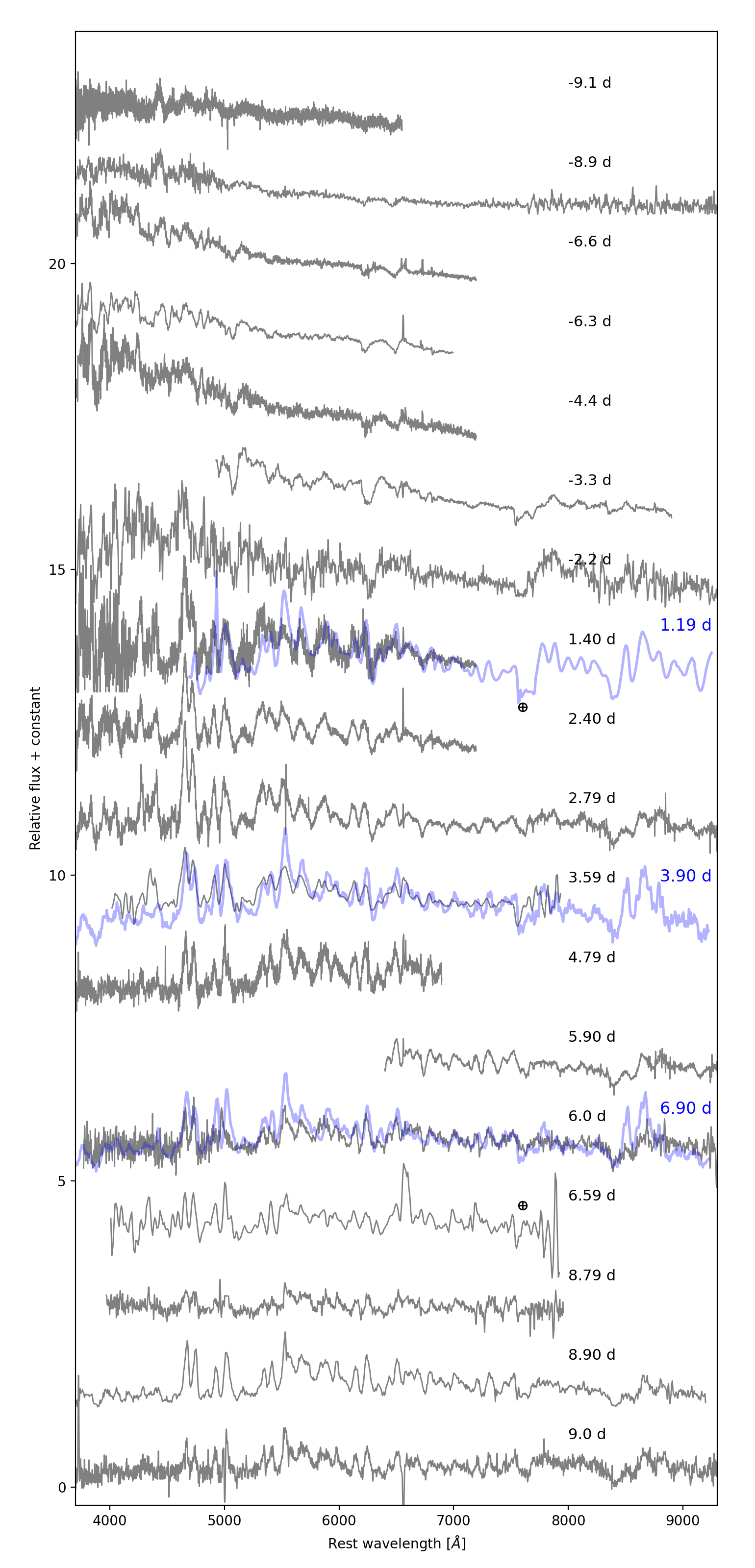}
    \caption{Optical spectra of SN 2022ywf. The epochs
show the days with respect to r-maximum. The observation log of spectra can be found in Tab. \ref{tab:22ywf_spectra}.  The first three spectra of SN 2023zgx (blue) close in phase to those of SN 2022ywf (grey) are also shown as comparison. }
    \label{fig:sn22ywf_spectroscopy}
\end{figure}

The ATLAS \textit{co}-band LCs provide early detections supplemented with deep non-detection limits ($> 20.7$ mag). The steeply rising early part of the SN LCs can be fitted with the so-called simple \textit{fireball} model \citep{Riess99}, where we assume a constant photospheric temperature ($T_\mathrm{eff}$) and velocity ($v_\mathrm{phot}$); thus, the emerged flux is a quadratic function of the time since the explosion ($F\, \sim \, t_\mathrm{exp}$). \citet{Firth15} showed on a large sample of normal Ia SNe that these assumptions are not robust, and while the early rise of the LC can be fitted with a power-law function:

\begin{equation}
F = a \cdot \left(t - t_{\mathrm{exp}}\right)^{n},
\label{eq:fireball}
\end{equation}

the $n$ exponent, referred to as rise index, covers a wide range between 1.5 and 2.5 and has a mean value of $n=2.44$. In the case of the subluminous Type Iax SNe, the fit of the rise index results in a typically lower value. \cite{Magee25} recently studied the rising curves of all available SNe Iax with detailed pre-maximum coverage and found that rise indices of SNe Iax show a similar scattering to normal SNe Ia, but with a lower mean value of 1.4-1.5 in the g- and r-bands. Their sample was mainly limited to the RL group but also included the faintest type Iax SN 2024vjm ($M_\mathrm{L} = -13.19 \pm 0.15$ mag), for which a rise index of $n_\mathrm{L} = 0.94^{+0.32}_{-0.34}$ was measured. Similar analyzes of individual SNe showed that the least luminous objects \citep[e.g. SNe 2015H, 2019muj and 2020udy,][respectively]{Magee16,Barna21,Singh24} have rise indices $n\, < \, 1.3$, indicating that there might be a link between luminosity and the rising parameters of type Iax SNe.

Handling the rise index as a free parameter in the fitting of the SN 2022ywf o-band LC (see Fig. \ref{fig:sn22ywf_atlas}), which includes the most pre-maximum observations, would return an extremely low exponent ($n=0.38 \pm 0.09$) and might be an unrealistic first light date ($T_\mathrm{0}= 59880.25 \pm 0.22$ MJD) coinciding with the time of the discovery (59880.17 MJD). This inconsistency is probably due to the effect of insufficient coverage of the rising phase, with six pre-maximum observations in $o$-band and even fewer with other filters. Thus, we assume the time of first light based on the last-non-detection and the first detection in $c$-band as $T_\mathrm{0}=59879.8 \pm 0.6$ MJD. 

As an alternative way to estimate $T_\mathrm{0}$, and thus the rise time, the abundance tomography analysis (see Sec. \ref{22ywf_spec_analysis}) also provides constraints on the time of first light, as the fit of the spectral time series includes the input parameter $t_\mathrm{exp}$ of the spectral fitting as the dilution factor of the density structure. However, the constrained densities mark the date of the explosion, which occurs earlier than the escape of the first photons from the ejecta, i.e., first light. The interim period, also called the dark phase, may cover a few days for SNe Ia \citep{Piro13} and explosion models with a compact $^{56}$Ni distribution \citep{Magee20}. On the other hand, deflagration models predict dark phases shorter than one-day \citep{Fink14}.

\begin{figure*}
\centering 
	\includegraphics[width=15cm]{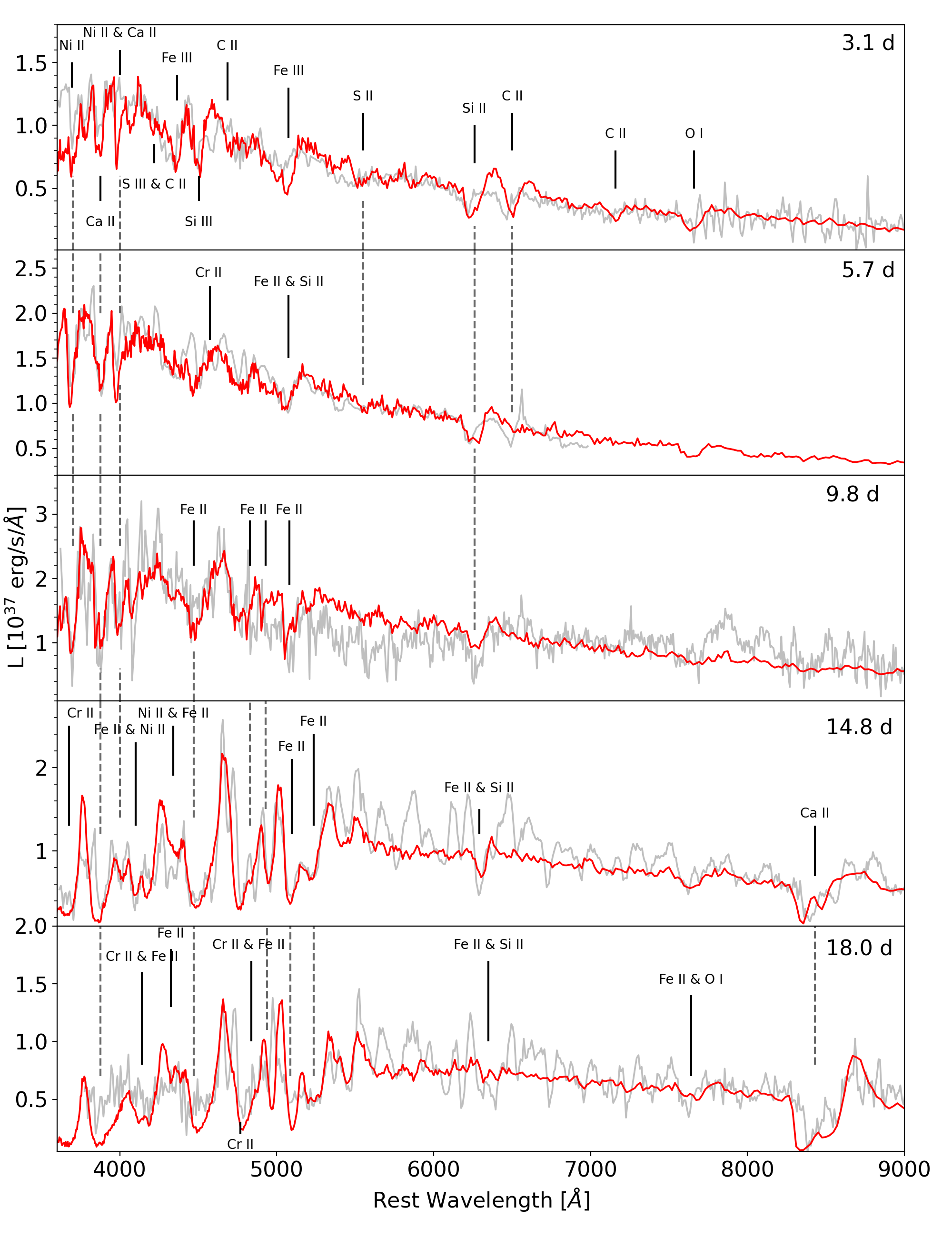}
    \caption{Five spectral epochs of SN 2022ywf (grey) fitted with the TARDIS synthetic spectra (red) produced in the abundance tomography analysis. }
    \label{fig:tardis_22ywf}
\end{figure*}

\begin{figure*}
\centering 
	\includegraphics[width=15cm]{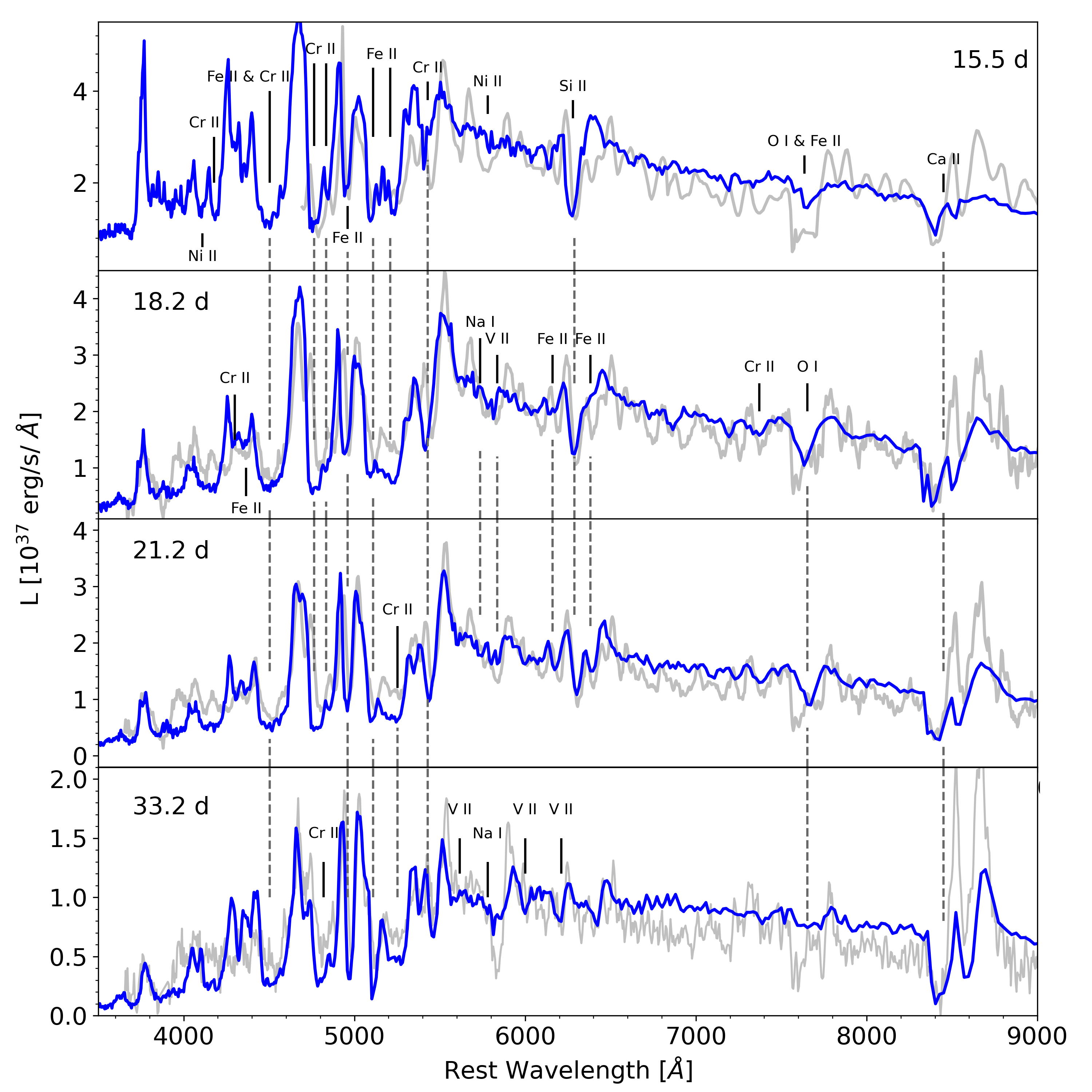}
    \caption{Four spectral epochs of SN 2023zgx (grey) fitted with the TARDIS synthetic spectra (blue) produced in the abundance tomography analysis. }
    \label{fig:tardis_23zgx}
\end{figure*}

SN 2023zgx, the other subject of this study, peaks around -14.4 mag in $r$-band, based on the similarities with SN 2022ywf in post-maximum LCs and spectral evolution (see Fig. \ref{fig:sn22ywf:photometry}), but the same set of photometric parameters cannot be precisely estimated due to the lack of coverage of the pre-maximum data. However, the many similarities between the two SNe suggest that we can safely assume a similar light curve evolution and constrain the time of peak brightness of SN 2023zgx based on that of SN 2022ywf. The declining LCs in g-, V- and r-bands can be overlapped with those of SN 2022ywf in the first two weeks after maximum by assuming a shift of $\Delta t=T_\mathrm{23zgx}(r_\mathrm{max}) - T_\mathrm{22ywf}(r_\mathrm{max}) = 400$ days along the time-axis (see Fig. \ref{fig:sn22ywf:photometry}) and $\Delta m_\mathrm{r}=+0.2$ mag along the magnitude-axis. Assuming the same $\Delta t$ provides an excellent match with the ATLAS c- and o-band light curves of the two SNe with a slightly greater shift in brightness ($\Delta m_\mathrm{o}$=0.5 mag). Assuming the presented match with the shifted LCs, we consider $T_\mathrm{23zgx}(r_\mathrm{max}) = 60290.4$ MJD as a rough estimate of the date of maximum, which is used as the reference time for epochs of SN 2023zgx hereafter, with an approximate uncertainty of $\pm 3$ days. The comparison of LCs may indicate that SN 2023zgx fades more slowly; however, it is important to note that this statement is mainly supported by the photometry after +20 days, when observations of SN 2022ywf become uncertain. The decline rates show significant differences in the B- and i-filters, which suggest different color evolutions for the two objects. The lack of pre-maximum photometry of SN 2023zgx limits the determination of the date of explosion. Quantitative estimates of $T_\mathrm{exp}$ are constrained by the abundance tomography analysis described in Sec. \ref{sec:spec_analysis}.

    \begin{table*}[!h]
      \caption[]{The light curve parameters of SN 2022ywf. }
      \centering
         \label{tab:phot_fits}
         \begin{tabular}{c | c c c c c} 
            \hline
            \noalign{\smallskip}
             & B & g & V & r & i \\
            \noalign{\smallskip}
            \hline
            \noalign{\smallskip}
            T$_\mathrm{max}$ [MJD] & 59885.6 (0.16) & 59886.6 (0.1) & 59687.2 (0.2) & 59890.4 (0.3) & 59891.3 (0.6) \\
            m [mag]& 18.18 (0.02)  & 18.12 (0.02) & 18.25 (0.02) & 18.06 (0.02) & 18.25 (0.04) \\
            M [mag] & -13.74 (0.44)  & -13.77 (0.44) & -13.68 (0.44) & -13.81 (0.44) & -13.59 (0.44)\\
            $\Delta$m$_\mathrm{15}$ [mag] & 1.65 (0.03) & 1.67 (0.03) & 1.19 (0.03) & 0.91 (0.04) & 0.58 (0.10)\\
            t$_\mathrm{rise}$ [days]& 5.6 (1.5) & 6.6 (1.6) & 7.6 (1.5) & 10.4 (1.6) & 11.3 (2.0) \\
            \noalign{\smallskip}
         \end{tabular} 
   \end{table*}

\section{Spectral analysis}
\label{sec:spec_analysis}

The optical spectra of SNe 2022ywf and 2023zgx are shown in Figs \ref{fig:sn22ywf_spectroscopy} and \ref{fig:sn23zgx_spectroscopy}, respectively. Due to the short sampling period of SN 2022ywf and the lack of pre-maximum epochs of SN 2023zgx, there is only a short period of a few days when the two spectral series overlap in phases. The three spectra of SN 2023zgx during this period show excellent matches with those of SN 2022ywf, which underlines the general similarity of the intrinsic properties of the two objects. SN 2022ywf evolves slightly faster, as can be seen in the shifting epochs of the matching spectra. This might be the result of the shorter diffusion timescale due to the lower ejecta mass, which can be tested by comparing the inferred density profiles from abundance tomography (see below).
 
 We analyze the first 20 days of the spectral time series of SNe 2022ywf and 2023zgx sampled with seven and four spectra, respectively, obtained at pre- and near-maximum epochs. In this period, the assumptions of TARDIS \citep{Kerzendorf14}, such as the existence of a sharp photosphere, are rather robust, allowing for the rapid modeling of realistic synthetic spectra for the optical regime. We aim to fit these spectral time series listed in Tabs. \ref{tab:22ywf_spectra} and \ref{tab:23zgx_spectra} follow the method of abundance tomography analysis. This technique assumes that line formation occurs mainly close to the photosphere; thus, each spectral epoch can constrain a certain layer of the ejecta. The retreating photosphere allows for mapping the ejecta inward by fitting the spectral time series accordingly. The later epochs may indicate a review of the outer layers as well, since some chemical elements could form lines in a more diluted but cooler region of the ejecta. 

 Abundance tomography of SNe Iax has been performed in multiple studies since the first adaptations \citep{Magee16, Barna17} of the TARDIS radiative transfer code \citep{Kerzendorf14}. The rapid evolution of the optical spectra, together with the numerous unblended spectral features that, compared to those of normal SNe Ia, are not saturated, provides suitable cases for fitting a large number of parameters of the ejecta through rapid modeling with a simplified radiative transfer code. SNe 2002cx, 2005hk, 2012Z \citep{Barna18}, 2019gsc \citep{Srivastav20}, 2019muj \citep{Barna21}, 2018cni, 2020kyg, \citep{Singh23} 2020udy \citep{Singh24} have been the subjects of similar abundance tomography, where the density profiles, abundance ratios, photospheric properties, and the times of explosions were constrained (the exact fitting strategy varies in each study; for further details, see the references). It is important to note that none of these modeling attempts aimed to find the best fit due to the large number of fitting parameters. Instead, our fitting process provides only a feasible solution for the model atmosphere (labeled as ''best-fit'' for simplicity) by testing and modifying the results of the \textit{N5def\_hybrid} \citep{Kromer15} pure deflagration model.
 

\subsection{SN 2022ywf}
\label{22ywf_spec_analysis}
\begin{figure}

	\includegraphics[width=\columnwidth]{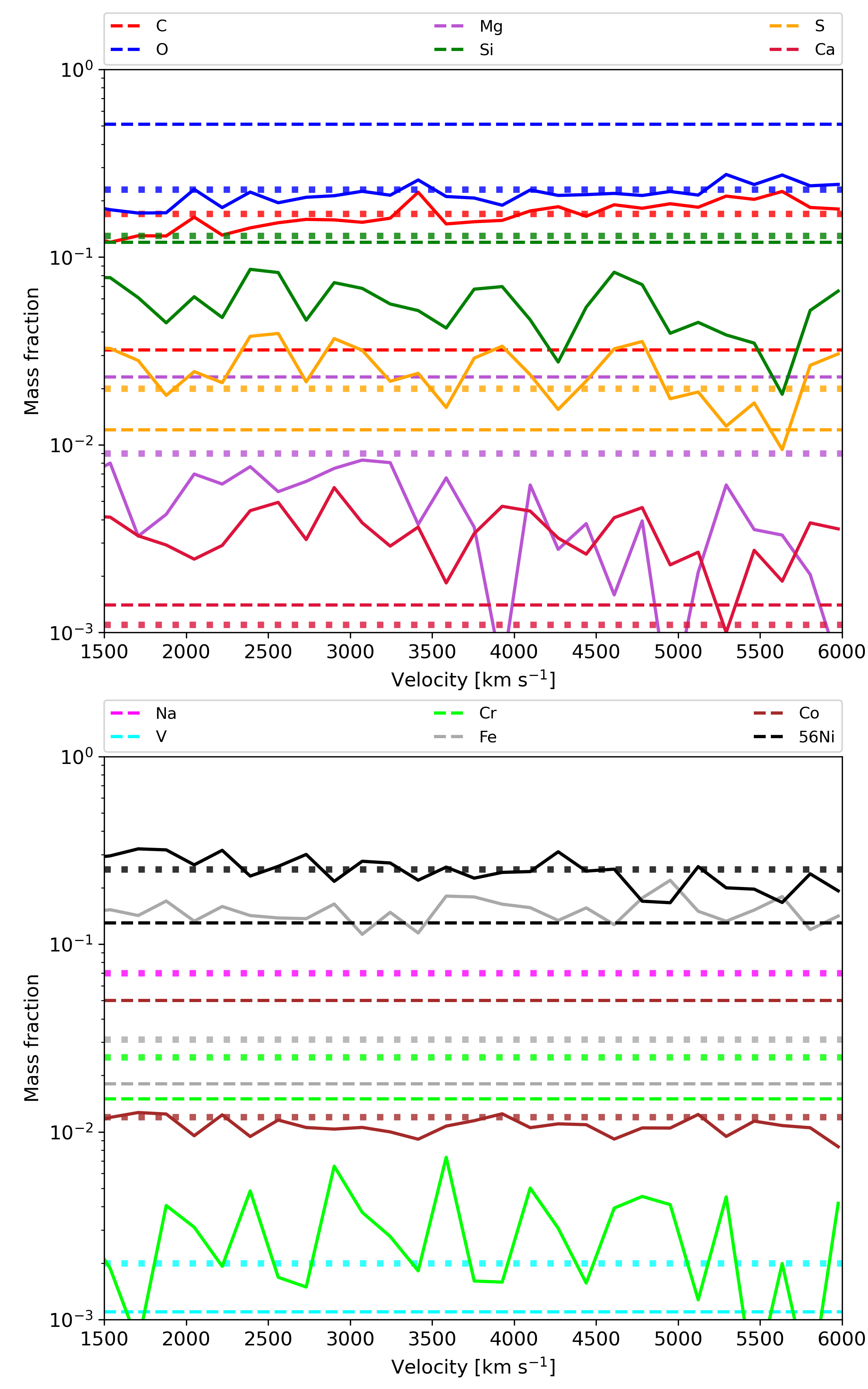}
    \caption{The mass fractions of chemical elements investigated in the abundance tomography analysis of SNe 2022ywf (dashed lines) and 2023zgx (dotted lines). The chemical abundances of the pure deflagration model \textit{N5def\_hybrid} (solid lines) are also shown for comparison. The mass fractions of the radioactive $^{56}$Ni are scaled to $t_\mathrm{exp}=100$s.} 
    \label{fig:abundances_22ywf}
\end{figure}

As a first approach, we reduce the number of free parameters by adopting constant abundances based on the predictions of a corresponding deflagration model, similar to the analysis of SN 2014dt described by \cite{Camacho-Neves23}. Taking the peak luminosities, the only hydrodynamic simulation with $M_\mathrm{V} \sim 14$ mag is \textit{N5def\_hybrid}, assuming the failed deflagration of a CONe-hybrid WD. The estimated observables of the hydro model showed a good match with the early photometric and spectroscopic evolution of the EL-Iax SN 2008ha, but it failed to reproduce the decline rate and the continuum flux weeks after the maximum. The adopted constant chemical composition based on the \textit{N5def\_hybrid} (see Fig. \ref{fig:abundances_22ywf}) is characterized by C, O, $^{56}$Ni, and stable IGE (mostly Fe) as the most abundant elements, each having mass fractions of around 20\%. While adopting this uniform chemical structure, we aim to fit the continuum and the line widths at each epoch, allowing us to constrain $\rho(v)$, $t_\mathrm{exp}$ and $v_\mathrm{phot}$, whose parameters affect the local temperature of the ejecta at first approximation.

For the density structure, we adopt a purely exponential function similar to that of N5def\_hybrid: 

\begin{equation}
    \rho(v,t_\mathrm{exp}) =  \rho_{0} \cdot exp^{-\frac{v}{v_{0}}} \cdot  \left( \frac{t_\mathrm{exp}}{t_0} \right)^{-3}
\label{eq:density_profile}
\end{equation}

where $\rho_{0}$ is the central density at $t_0$ reference time (typically chosen as $t_\mathrm{exp} =100$ s, the approximate end of the fusion processes and the beginning of homologous expansion). The $v_0$ parameter characterizes the slope of the density profile. This model does not exhibit any cut-off in the outer layers, such as the similarly constrained density profiles of RL SNe \citep{Barna18} and the corresponding hydrodynamic models \citep{Fink14}. Since the studied spectral series does not cover the vast majority of the ejecta, we cannot constrain the M$_\mathrm{ej}$ ejecta mass from abundance tomography; only the bolometric light curve fitting can provide an estimate of the ejecta mass. However, the fitted parameters of Eq. \ref{eq:density_profile} allow a quantitative comparison between the outer layers of other SNe Iax and deflagration models.

\begin{figure}

	\includegraphics[width=\columnwidth]{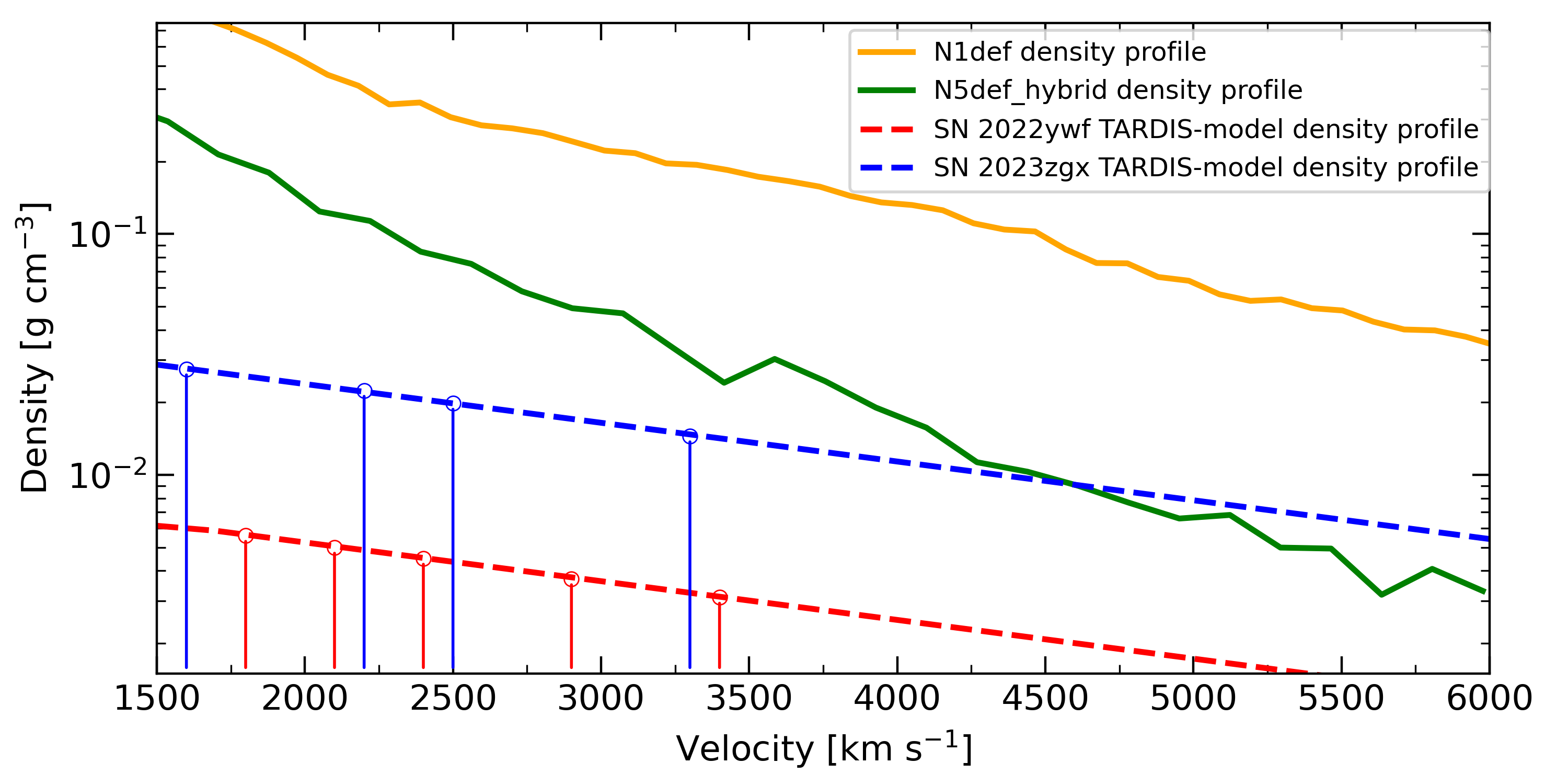}
    \caption{The exponential functions assumed for the density profiles of SNe 2022ywf (red) and 2023zgx (blue) in the best ejecta models, scaled to $t_\mathrm{exp}=100$ s. The density structures of the two least-luminous pure deflagration models are also plotted for comparison. The vertical lines indicate the location of the photosphere at the fitted epochs in the corresponding models. }
    \label{fig:densities}
\end{figure}

The fit of the density structure is relatively uncertain, as the two main properties defining the function, the $\rho_\mathrm{0}$ central density parameter and the $v_\mathrm{0}$ slope of the exponential function, are strongly coupled. The segment of the density function sampled by a certain spectral epoch mainly affects the optical depth of the lines, but it also influences the temperature profile of the ejecta, allowing us to constrain the densities. During the fitting process, we observed that TARDIS modeling is sensitive to changes in densities of at least 50\% as it caused discrepancies in the observed spectral features that could not be compensated for by modifying chemical abundances or other physical parameters. The density function used for the SN 2022ywf model (see in Fig. \ref{fig:densities}).

Compared to the other two fitting parameters in the density function, the date of the explosion can be fitted precisely since it affects the entire spectral time series via the dilution of the density profile (see $t_\mathrm{exp}$ in Eq. \ref{eq:density_profile}). The best-fit value is found to be $T_\mathrm{exp}=59878.4$ MJD. The early epochs, especially the first three between t$_\mathrm{exp} = 3.1$ and 5.7 days, are very sensitive to the $T_\mathrm{exp}$ parameter, as even a few hours can strongly modify the temperature profile and, thus, the continuum. Even though our method is not suitable for detailed uncertainty analysis, the estimated upper limit of the error of $T_\mathrm{exp}$ is $\pm 0.5$ day. This provides a dark phase between the first light and the explosion as $T_\mathrm{0} - T_\mathrm{exp} = 1.4$ days.

As a second step, we continue the spectral fitting by allowing the constant mass fractions of elements that have a significant impact on the spectral features, namely C, Mg, Si, S, Ca, Fe, and the most abundant radioactive isotope, $^{56}$Ni, to vary. Note that TARDIS uses the actual state of decay products at each epoch, thus, abundances of Co and Fe increase with time. Other elements and isotopes, such as Ne, although relatively abundant ($X(Ne) \simeq 0.03$), have a limited impact on the early optical evolution, and we keep their initial mass fraction fixed. Oxygen is used as a buffer element to manage the normalization of the total mass fraction of chemical elements to 1. Furthermore, the features of post-maximum epochs indicate the inclusion of Cr and V, two elements that may originate from the radioactive decay of $^{54}$Fe but are expected to be present in low abundances. The required mass fractions ($X(Cr) \simeq 0.01$; $X(V) \simeq 0.001$) are added to the models of each epoch to maintain the concept of a uniform chemical structure despite the slight deterioration of the fitting in the early epochs (i.e., in the outer region of the ejecta). This behavior implies the stratified distribution of these elements.

The best-fit ejecta model contains solely constant chemical mass fractions similar to the abundance structure of the $N5def\_hybrid$ model used for the initial input. This simplification allowed us to handle the abundances of the aforementioned elements as free parameters. The synthetic spectra of our best-fit model successfully reproduce all prominent features throughout the studied three week time span (see Fig. \ref{fig:tardis_22ywf}). Inconsistencies, such as the shift of certain spectral lines, arise due to the simplifications of the fitting method and/or the limitations of the radiative transfer code.

 Such mismatches are the shifted synthetic lines that may occur due to the incorrect assumption of constant abundances. The C II $\lambda$6580 lines show a consistent redshift compared to the observations, indicating that some level of stratification may be necessary for carbon in the outer layers of the ejecta. Another shortcoming of the synthetic spectra is the lack of several features between 6000 and 8300 \r{A} after maximum, observed as narrow P Cygni lines and previously identified as C II and Fe II features \citep{Tomasella16}. Since the strong iron features formed close to the photosphere are relatively well reproduced at $t_\mathrm{exp}=14.8$ and 18.0 days, and both $X(C)$ and $X(Fe)$ have constant values throughout the model atmosphere, this issue probably originates from the outer, cooler region of the model ejecta.
 
The Si and S mass fractions are modified by a few percentage points to improve the fits of the characteristic spectral features $\lambda6345$ and $\lambda5607$, respectively. The fractions of two low-abundance elements, Ca and Cr, are changed by factors of $\sim$3 to fit the short wavelength end of the spectra as well as the characteristic post-maximum spectral features Cr II $\lambda5276$ and the Ca II NIR triplet. However, these discrepancies in the $N5def\_hybrid$ hydro model are not significant, as the changes are only up to a few tenths of a percentage point.
The main differences arise in IGE abundances, as both the radioactive Ni and the stable Fe are present with significantly lower mass fractions in our model. These updates are required by the fit of the two earliest spectra to reproduce the corresponding lines (see Fig. \ref{fig:tardis_22ywf}). However, even the reduction of both $X(Fe)$ and $X(^{56}Ni)$ mass fractions by $\sim$0.12 produces excessively strong Fe III $\lambda4420$, $\lambda5157$, and Ni II $\lambda4068$ lines for the spectrum obtained at $t_\mathrm{exp}= 3.1$ days, but provides a relatively good match for the spectral features at $t_\mathrm{exp}= 5.7$ and 9.8 days. This inconsistency is caused by the use of a uniform chemical structure, as the outer ejecta of SNe Iax may have even lower IGE abundances.

   \begin{table}[h]
      \caption[]{Physical parameters of the best-fit TARDIS models for SNe 2022ywf and 2023zgx.  }
         \label{tab:tardis_fits}
         \begin{tabular}{c | c | c} 
            \hline
            \noalign{\smallskip}
             & SN 2022ywf & SN 2023zgx \\
            \noalign{\smallskip}
            \hline
            \noalign{\smallskip}
            $\rho_0$ [g\,cm$^{-3}$] & \multirow{2}{*}{0.01} &  \multirow{2}{*}{0.05} \\
            central density & & \\
            \noalign{\smallskip}
            \hline
            \noalign{\smallskip}
            $v_0$ [km\,s$^{-1}$] & \multirow{3}{*}{2500} &  \multirow{3}{*}{2700} \\
            exponential factor & & \\
            of the density profile & & \\
            \noalign{\smallskip}
            \hline
            \noalign{\smallskip}
            $v_\mathrm{phot,m}$ [km\,s$^{-1}$] & \multirow{3}{*}{2400} & \multirow{3}{*}{3100} \\
            photospheric velocity & & \\
            at $T_\mathrm{max}(r)$ & & \\
            \noalign{\smallskip}
            \hline
            \noalign{\smallskip}
            $v_\mathrm{phot,15}$ [km\,s$^{-1}$] & \multirow{3}{*}{2200} & \multirow{3}{*}{3100} \\
            photospheric velocity & & \\
            at $t_\mathrm{exp}=15$ days & & \\
            \noalign{\smallskip}
            \hline
         \end{tabular} 
   \end{table}

\subsection{SN 2023zgx}
\label{23zgx_tomography}

The spectral series of SN 2023zgx covers a significantly longer period, almost three months of spectral evolution. Although type Iax SNe never go fully nebular, and thus the radiative transfer code assuming a sharp photosphere can be used to reproduce part of the optical regime \citep{Camacho-Neves23}, the emerging forbidden lines place limitations on the quality of fitting. Because of this, and due to the comparison with the fitting of SN 2022ywf, we fit only the first four epochs (the first five weeks after the explosion, approximately) of the spectral series. The fitting process was repeated with the same principles; however, in contrast to the case of SN 2022ywf, fewer fitting constraints were available due to the lack of early spectra sampling the outermost regions of the ejecta. This circumstance leaves the physical parameters even more uncertain, especially regarding layers over 5000 km\,s$^{-1}$, which are already at a distance from the photosphere at the first epoch. 

By fitting only the post-maximum epochs of SN 2023zgx, its chemical composition is less constrained. This is partially because synthetic spectra, in general, become less sensitive to changes in mass fractions over time. Moreover, the weaker lines cannot be well constrained due to the presence of the strong and overlapping Fe II features. Thus, the distribution of chemical elements that have a significant impact on the pre-maximum spectral features, such as C, S, and Ni, can be effectively described with the mass fraction obtained from the $N5def\_hybrid$ model, which is used as initial input.

Despite the later triggering of follow-up observations, the tomography analysis investigated the same velocity coordinates as those of SN 2022ywf due to its lower expansion velocities. The slope of the inferred ejecta structure of the two SNe is similar, but the layers at the same velocities are denser in the model of SN 2023zgx to reproduce the strong post-maximum IGE features. This may also explain the higher photospheric velocities of SN 2023zgx at the same epochs, regardless of whether they are relative to the maximum or the moment of explosion. This relation also fits into the trend of deflagration models, where more energetic (and luminous) explosions produce more massive ejecta having higher densities, even in the outer regions \citep{Fink14,Lach22} The angle-averaged density function of N5def\_hybrid matches relatively well with that of the best SN 2023zgx model above 3000 km\,s$^{-1}$, but strongly exceeds it in the deeper ejecta.

The time of the explosion is also more uncertain since $t_\mathrm{exp}$ mainly impacts the spectrum through the dilution of the densities. While the date of the explosion can be narrowed down to approximately $\pm1$ day precision in the case of fitting early epochs, the synthetic spectra of SN 2023zgx are relatively insensitive to a change of $\sim2$ days in $t_\mathrm{exp}$ due to the lack of pre-maximum spectra. The tomography analysis constrained the date of the explosion to $T_\mathrm{exp}=60276.1$ MJD, which provides a similar rise time to that of SN 2022ywf.

The synthetic spectra provided by the best-fit model (see Fig. \ref{fig:tardis_23zgx}) show a good match with the observed ones, as almost all prominent lines are present with the proper optical depth. We also added a high Na abundance to the model ejecta to fit the growing feature around 5800 \r{A}, which has become one of the most prominent P Cygni features at $t_\mathrm{exp}=33.2$ days, similarly to the late-time spectral synthesis of SNe 2014dt and 2022xlp \citep[][respectively]{Camacho-Neves23,Banhidi25}. The only striking difference concerns the NIR triplet of Ca II, where the rising emission of forbidden lines disrupts the fitting based on the photospheric assumptions. Another shortcoming of the fit is that the Fe II lines are too strong between 4200 and 5400 \r{A}, resulting from Fe in the outer regions of our model ejecta. The flux continuum is also reproduced for all four spectra, indicating reasonable estimates for the photospheric velocity and the density profile (including $t_\mathrm{exp}$). Comparing the fits to those of SN 2022ywf, we can conclude that the spectral evolution of the post-maximum epochs can be better reproduced by the adopted assumptions, indicating a uniform abundance structure for the inner regions of the ejecta.


The constrained ejecta parameters of the best-fit TARDIS models are listed in Tab. \ref{tab:tardis_fits}. The density structures and chemical abundances of SN 2022ywf are shown in Figs. \ref{fig:densities} and \ref{fig:abundances_22ywf}, along with those of SN 2023zgx. We note that our fitting strategy aimed to test the predictions of the corresponding hydrodynamic model \textit{N5def\_hybrid}; thus, the differences between the ejecta models of SNe 2022ywf and 2023zgx are not necessarily real. The constant chemical structure was obtained based on the N5def\_hybrid model with modified iron ($X(Fe)=0.025$), chromium ($X(Cr)=0.01$) and sodium ($X(Na)=0.07$)abundances. These changes are mostly consistent with the constrained model atmosphere of SN 2022ywf. Moreover, the abundance structure of SN 2022ywf would result in an indistinguishably good fit for SN 2023zgx as well. This means that those abundances, which differ from the \textit{N5def\_hybrid} model to fit the early spectral features of SN 2022ywf (e.g., decreased carbon and nickel mass fractions), have no strong impact on spectral formation at post-maximum epochs.

\subsection{Comparison with other extremely faint SNe Iax}

So far, only five SNe Iax from the faintest end of the luminosity range of the subclass have been published (i.e. with $M_V \, < \, 14$ mag) with detailed follow-up observations before and around maximum light. In this section, we compare the optical spectral evolution and features of SNe 2022ywf and 2023zgx with those of the faint Iax objects, namely SNe 2008ha \citep{Foley09}, 2020kyg \citep{Singh24}, 2019gsc \citep{Srivastav20, Tomasella20} and 2021fcg \citep{Karambelkar21}.

\begin{figure}

	\includegraphics[width=\columnwidth]{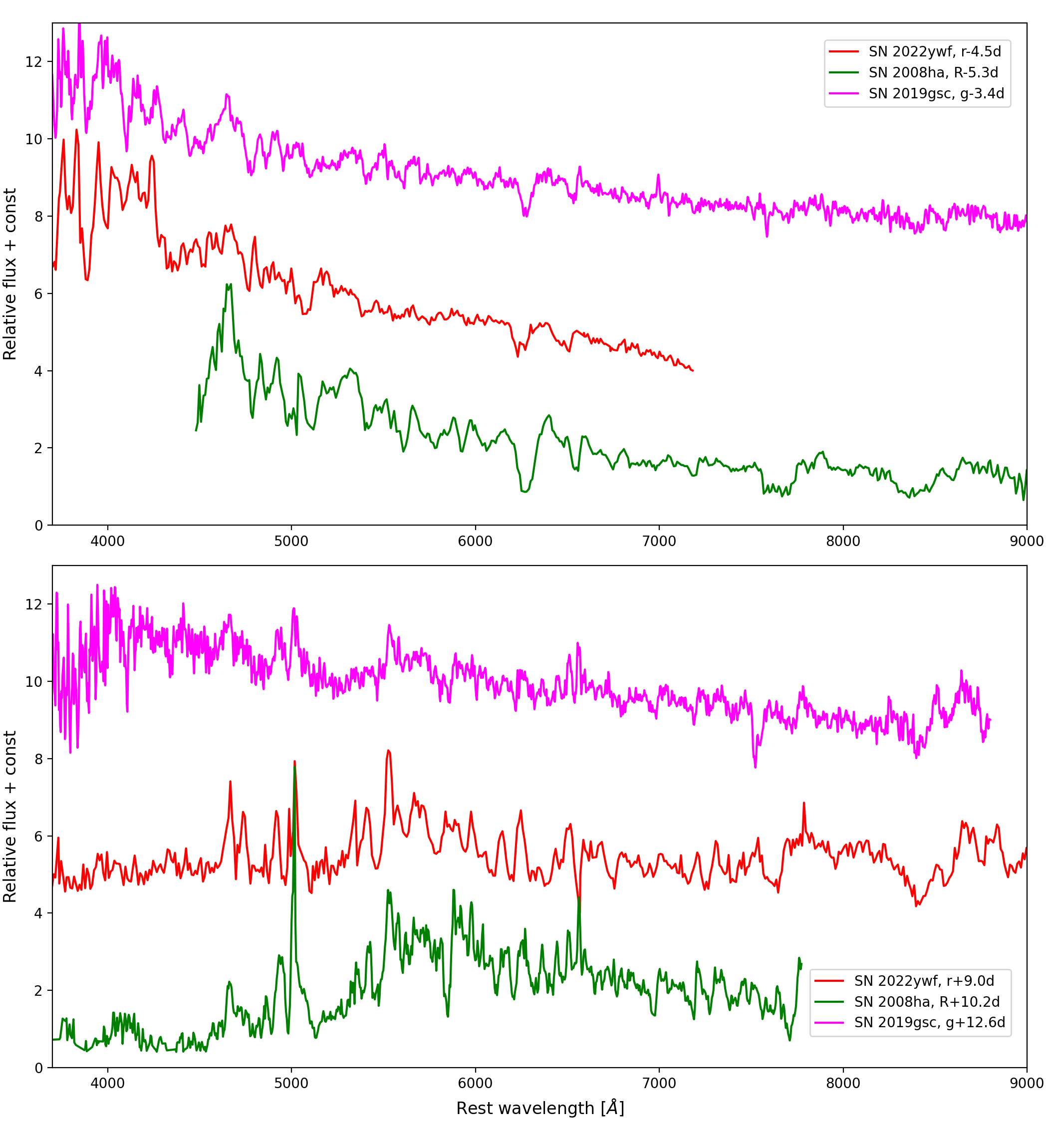}
    \caption{Spectral comparison at similar epochs between three SNe Iax from the EL sample. The spectra are corrected for reddening and redshift; epochs are taken from the corresponding publications of SNe 2008ha \citep{Foley09, Valenti09} and 2019gsc \citep{Srivastav20}.}
    \label{fig:spec_comps}
\end{figure}

In Fig. \ref{fig:spec_comps}, we compare the corresponding spectral epochs before the r-band maximum and approximately 10 days after that. Even though the EL sample shares some common features in their spectral evolution, such as the presence of the Si II $\lambda 6355$ and the C II $\lambda6580$ lines, the strength of these spectral features greatly differs for objects with similar luminosities. These differences are amplified for the post-maximum epoch, when SN 2008ha shows wide absorption of IGEs, while SN 2022ywf represents a more typical set and appearance of P Cygni lines, resembling those of SN 2005hk. SN 2019gsc, on the other hand, lacks strong (pseudo-)emission peaks and shows a significantly slower spectral evolution over 16 days. Since the identified lines are similar, the variation in spectral features may be caused by the different ejecta temperatures. This idea is supported by the color and, thus, the continuum evolution of the objects compared here. While the ejecta temperature of SN 2019gsc barely changes between the epochs in this comparison, SN 2022ywf shifts from a hotter ejecta
to a cooler phase, changing the initially presented Fe III and Ni III features to a Fe II dominated post-maximum epoch. The faster cooling of SN 2022ywf can be linked to a steeper density gradient, which is caused by either the different ejecta mass or a line of sight effect, where we have insight to a less dense region of the SN.

The correlation between the peak luminosities and other observables of SNe Iax have been investigated by several studies. \citet{Singh23} and \citet{Magee25} presented a correlation between the rise time ($t_\mathrm{rise}$) and the peak absolute magnitude in the r-band, as brighter SNe Iax typically reach their maximum later. However, the demonstration of the correlation is still ambiguous due to the limited sample and the high uncertainty of the estimated $t_\mathrm{rise}$ parameters resulting from the relatively late discoveries. The decline parameter $\Delta$m$_{15}$, which shows a greater variance, rather indicates clustering, where the brighter and fainter samples seem to show different behavior \citep{Singh23}. The lack of strong correlations and the continuous distribution of physical properties throughout the whole luminosity range may indicate that not all transients labeled as type Iax SNe originate from the same explosion scenarios. 

Another possible correlation between peak luminosity and expansion velocity has been referred to by multiple studies. \citet{Tomasella16} investigated a small sample of SNe Iax with spectral coverage around their maximum and constrained $v_\mathrm{phot}$ values. The authors found no proof that SNe Iax with higher velocity also have lower peak absolute magnitudes, as the trend of five objects is spoiled by two outliers, SNe 2009qu \citep{Narayan11} and 2014ck \citep{Tomasella16}, with higher than expected velocities. We note that these statistics are based on sometimes inconsistently estimated $v_\mathrm{phot}$ values. The standard method of calculating the current location of the photosphere, as it is adopted from the analysis of normal SNe Ia, is based on the blueshift of the absorption feature of the prominent Si II line. Multiple studies \citep{Szalai15, Maguire23} showed that this method is not reliable for SNe Iax, as the $\lambda6355$ feature overlaps with an Fe II absorption comparable even at pre-maximum epochs.
\citet{Barna21} reported a stronger trend between peak luminosity and $v_\mathrm{phot}$ estimated at the moment of B-band maximum light, when the velocities are constrained by abundance tomography. The tomography analysis carried out with a radiative transfer code not only supports the consistency in the comparison of various SNe Iax, but the method also fits the continuum flux and various spectral lines together. This allows us to constrain the bottom of the line forming regions simultaneously and to avoid the incorrect fits due to the misidentification of lines, providing a more certain estimate of v$_\mathrm{phot}$. 

\begin{figure}

	\includegraphics[width=\columnwidth]{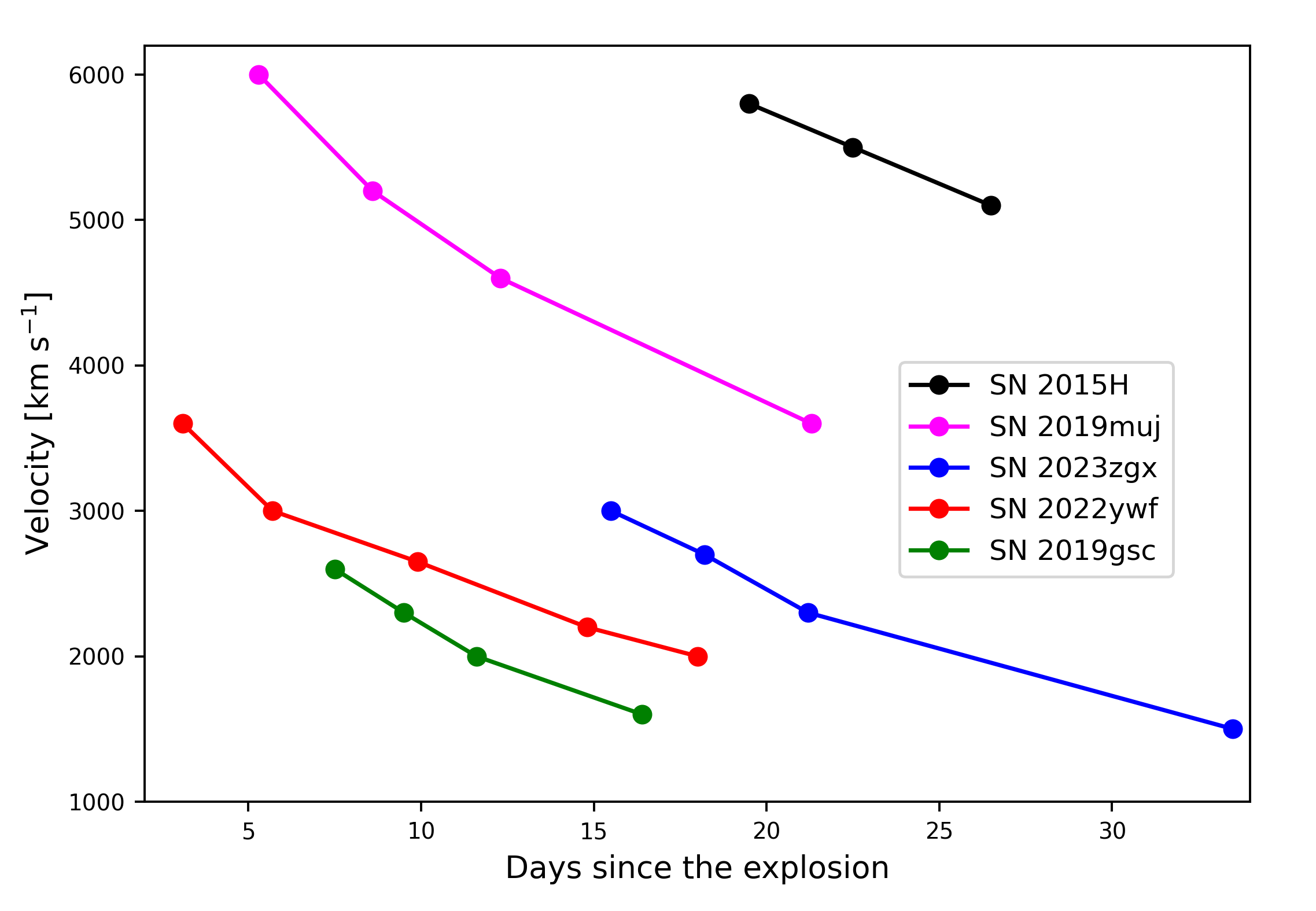}
    \caption{Photospheric velocity evolution of EL and IL SNe Iax constrained by abundance tomography method. }
    \label{fig:velocity}
\end{figure}

\begin{figure}

	\includegraphics[width=\columnwidth]{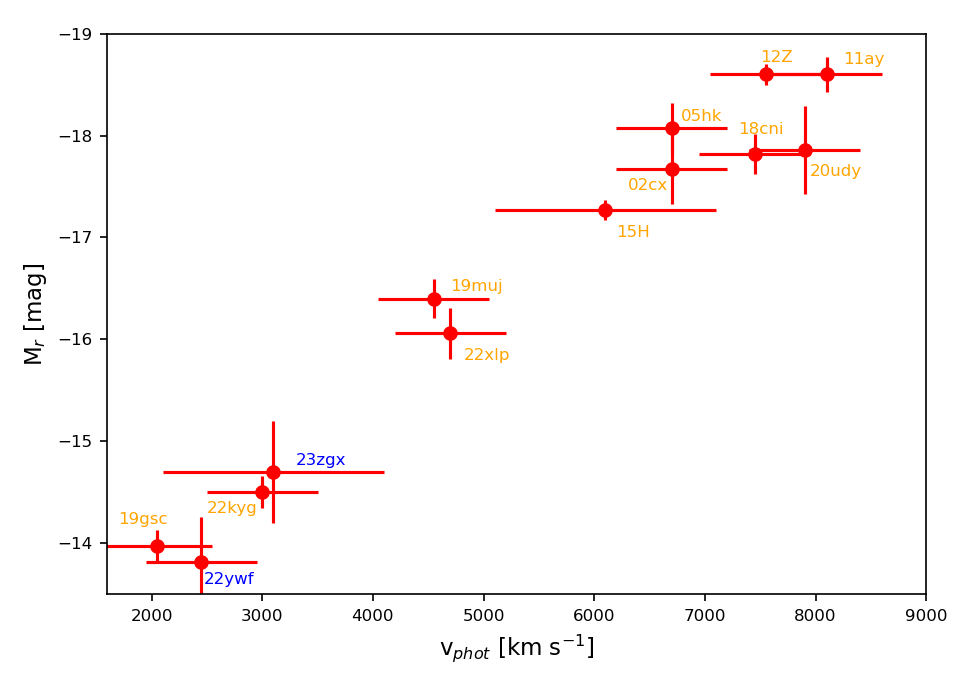}
    \caption{Absolute magnitudes in r/R-band of SNe Iax as a function of their photospheric velocity at the moment of maximum light.  We only plot those SNe Iax which have been the subject of abundance tomography. In case of no spectral coverage of the maximum (SNe 2015H and 2023zgx), photospheric velocity is estimated with linear extrapolation. The corresponding references are listed in Tab. \ref{tab:citations}.}
    \label{fig:vM}
\end{figure}

In Fig. \ref{fig:velocity}, the evolution of v$_\mathrm{phot}$ of SNe 2022ywf and 2023zgx are matched with that of other EL SNe Iax. The covered v$_\mathrm{phot}$ curves plotted as a function of t$_\mathrm{exp}$ do not cross each other, which supports the assumption that the relationship between peak luminosity and expansion velocity extends to the extremely low-luminosity range of the subclass as well. For better comparison, we also illustrate the relation of the two physical properties for all SNe Iax with well-studied spectral series in Fig. \ref{fig:vM}, where the v$_\mathrm{phot}$ values are obtained at the moment of r-band maximum. We include only those cases where the velocity evolution was mapped by spectral synthesis of radiative transfer codes and had coverage around the maximum light. Assuming monotonically decreasing v$_\mathrm{phot}$ for these epochs, we interpolate to calculate the velocity value for the moment of r- (or, if missing, R-)band maximum. These conditions limit the sample to 13 SNe, which cover almost the entire luminosity range of the subclass, with gaps in the IL group. The seemingly strong correlation between peak luminosities and expansion velocities further supports the assumption that all SNe Iax, including RL, IL, and EL, share a similar origin. Furthermore, the shown trend is not consistent with the scenario of a WD merger resulting in failed detonation, as the simulation predicts relatively high photospheric velocities in the case of an even less luminous SN \citep[$M_\mathrm{V}=-11.3$ mag, $v_\mathrm{phot}=3900$ km\,s$^{-1}$][]{Kashyap18}. Note that the estimation of v$_\mathrm{phot}$ values in Fig. \ref{fig:vM} also required distances and is not fully independent of the constrained luminosities. Thus, the assumed correlation requires further investigation.

\section{Conclusions}
\label{sec:conclusions}

We have presented the photometric and spectroscopic dataset of two SNe Iax from the extremely low-luminosity (EL) regime of the subclass. SN 2022ywf has been the target of the most detailed spectroscopic follow-up from the EL sample so far, including multiple pre-maximum epochs; however, the monitoring ended two weeks after the r-band maximum. The observations of SN 2023zgx started only after its peak, but the follow-up covers $\sim100$ days. In this study, we mainly focused on the first month after the explosions, when the photospheric assumption is relatively robust for SNe Iax, and the observations of the two objects overlap. 

The comparison of the spectroscopic datasets showed that the first three epochs of SN 2023zgx (between +1.4 and +7.1 days) almost perfectly fit the spectra of SN 2022ywf. However, the corresponding epochs indicate a slightly faster evolution for SN 2022ywf, which may be linked to the steeper density gradient of the ejecta. Similar conclusions can be drawn from the comparison of other EL SNe Iax with similar spectral epochs.

The spectroscopic time series of SNe 2022ywf and 2023zgx were studied by the abundance tomography method, which allowed us to map the ejecta properties of both objects.
   \begin{enumerate}
      \item The comparison of the observables with other objects of the EL sample revealed  differences in the post-maximum LC and color evolution. 
      \item The density structures inferred from the abundance tomography analysis of SNe 2022ywf and 2023zgx are comparable to those of the $N5def\_hybrid$ model and, thus, to each other. Considering that both SNe have lower peak luminosities than the pure deflagration model, the lower densities are consistent with expectations.
      \item The derived uniform abundance models are also consistent with the predictions of the deflagration models but show slightly lower IGE mass fractions.
      \item The photospheric evolution of the two studied SNe fits into the general trend of type Iax SNe, as the more luminous objects tend to show higher velocities at the same epochs. 
   \end{enumerate}

The analysis of SNe 2022ywf and 2023zgx revealed that the evolution of the extremely low-luminosity objects can be efficiently described by the general properties of the pure deflagration models, such as the constant chemical abundances and a steep density profile. These results support the assumption that, despite the diversity in observables, the EL Iax SNe share the same origin. In the wider context, the intrinsic similarities can be extended to all SNe Iax. We showed that the correlation between peak absolute magnitudes and expansion velocities at the time of maximum light may extend to the extremely low-luminosity group. The continuous distribution of certain physical properties through the luminosity range of the subclass makes it probable that all SNe Iax originate from similar explosion scenarios.

\begin{acknowledgements}

This work uses data from the Las Cumbres Observatory Global Telescope network. The LCO team is supported by NSF grants AST-2308113 and AST-1911151. Based on observations collected at the European Organisation for Astronomical Research in the Southern Hemisphere, Chile, as part of ePESSTO+ (the advanced Public ESO Spectroscopic Survey for Transient Objects Survey – PI: Inserra). ePESSTO+ observations were obtained under ESO program IDs 108.220C and 112.25JQ. This project has received funding from the HUN-REN Hungarian Research Network. 
B.B. received support from the Hungarian National Research, Development and Innovation Office grants OTKA PD-147091. T.-W.C. acknowledges the financial support from the Yushan Fellow Program by the Ministry of Education, Taiwan (MOE-111-YSFMS-0008-001-P1) and the National Science and Technology Council, Taiwan (NSTC grant 114-2112-M-008-021-MY3). T.E.M.B. is funded by Horizon Europe ERC grant no. 101125877. Time-domain research by the University of Arizona team and D.J.S. is supported by National Science Foundation (NSF) grants 2108032, 2308181, 2407566, and 2432036 and the Heising-Simons Foundation under grant \#2020-1864. TP acknowledges the financial support from the Slovenian Research Agency (grants I0-0033, P1-0031, J1-8136, J1-2460 and Z1-1853). KAB  is supported by an LSST-DA Catalyst Fellowship; this publication was thus made possible through the support of Grant 62192 from the John Templeton Foundation to LSST-DA. Supernova research at Rutgers University is supported in part by NSF grant AST-2407567. SWJ also gratefully acknowledges support from a Guggenheim Fellowship. KAB is supported by an LSST-DA Catalyst Fellowship; this publication was thus made possible through the support of Grant 62192 from the John Templeton Foundation to LSST-DA.

\end{acknowledgements}

\bibliographystyle{aa}
\bibliography{aanda}

\onecolumn
\appendix
\section{Some extra material}

\begin{center}
\begin{longtable}{lccc}
\caption{Log of ATLAS photometry of SN 2022ywf.}\\
\label{tab:ATLAS_22ywf_phot}
\textbf{MJD} & \textbf{Filter} &\textbf{Flux} & \textbf{mag} \\ \hline \hline
59875.18 & c & 9.0 (5.97) & >20.77 \\
59876.42 & c & 2.6 (6.34) & >20.70 \\
59876.95 & o & 31.1 (11.8) & >20.03 \\ 
59878.41 & o & 3.21 (6.34) & >20.70 \\
59879.19 & c & 10.9 (6.34) & >20.70 \\
59880.45 & c & 52.7 (6.71) & 19.60 (0.14) \\ 
59880.77 & o & 56.2 (11.76) & 19.53 (0.23) \\ 
59882.41 & o & 114.1 (5.97) & 18.76 (0.06) \\ 
59883.16 & o & 140.5 (5.87) & 18.53 (0.05) \\ 
59884.41 & o & 164.1 (6.84) & 18.36 (0.05) \\ 
59884.90 & o & 156.9 (23.1) & 18.41 (0.16) \\ 
59886.36 & o & 149.0 (7.88) & 18.47 (0.06) \\ 
59887.11 & o & 174.3 (13.4) & 18.30 (0.08) \\ 
59892.48 & o & 202.4 (15.2) & 18.13 (0.08) \\ 
59892.92 & o & 171.2 (17.8) & 18.32 (0.11) \\ 
59894.38 & o & 176.0 (8.2) & 18.29 (0.05) \\ 
59896.45 & o & 172.3 (9.2) & 18.31 (0.06) \\ 
59896.88 & c & 103.4 (5.7) & 18.86 (0.06) \\ 
59898.39 & c & 120.0 (3.8) & 18.70 (0.04) \\ 
59899.12 & o & 148.0 (7.0) & 18.47 (0.05) \\ 
59900.49 & o & 141.4 (13.7) & 18.52 (0.11) \\ 
59900.86 & c & 82.4 (7.2) & 19.11 (0.10) \\ 
59902.30 & c & 57.5 (3.2) & 19.50 (0.06) \\ 
59903.11 & o & 118.8 (6.6) & 18.71 (0.06) \\ 
59906.34 & c & 65.7 (4.6) & 19.36 (0.08) \\ 
59907.10 & o & 114.9 (9.5) & 18.75 (0.09) \\ 
59908.35 & o & 91.5 (7.6) & 19.00 (0.09) \\ 
59908.97 & c & 61.8 (7.0) & 19.42 (0.12) \\ 
59911.09 & o & 83.7 (8.3) & 19.09 (0.11) \\
59912.84 & o & 65.6 (7.3) & 19.36 (0.12) \\
59915.05 & o & 77.4 (15.0) & 19.18 (0.21) \\
59923.11 & o & 46.6 (9.8) & 19.73 (0.23) \\
59927.09 & c & 25.0 (4.7) & 20.41 (0.21) \\
59928.31 & c & 32.2 (5.4) & 20.13 (0.18) \\
59931.06 & c & 29.8 (5.3) & 20.22 (0.19) \\
		\hline
\end{longtable}
\end{center}
\newpage

\begin{center}
\begin{longtable}{lccc}
\caption{Log of ATLAS photometry of SN 2023zgx}\\
\label{tab:ATLAS_23zgx_phot}
\textbf{MJD} & \textbf{Filter} & \textbf{Flux} & \textbf{mag} \\ \hline \hline
60294.63 & o & 351.3 (8.4) & 17.53 (0.02)\\
60296.61 & c & 230.4 (10.3) & 17.99 (0.04)\\
60299.34 & o & 290.0 (10.6) & 17.74 (0.04)\\
60301.07 & o & 244.6 (9.03) & 17.92 (0.04)\\
60304.63 & o & 188.1 (10.3) & 18.21 (0.05)\\
60306.57 & o & 184.2 (11.0) & 18.23 (0.06)\\
60308.59 & o & 164.6 (9.9) & 18.35 (0.06)\\
60309.54 & o & 149.3 (12.0) & 18.46 (0.08)\\
60310.50 & o & 132.2 (17.1) & 18.59 (0.14)\\
60314.61 & o & 127.9 (7.5) & 18.63 (0.06)\\
60324.59 & o & 97.6 (6.4) & 18.92 (0.07)\\
60327.35 & o & 78.2 (9.6) & 19.16 (0.13)\\
60329.02 & c & 58.0 (5.9) & 19.49 (0.11)\\
60331.33 & o & 94.9 (6.4) & 18.95 (0.07)\\
60333.04 & o & 66.0 (12.9) & 19.35 (0.21)\\
60336.51 & o & 71.3 (12.6) & 19.26 (0.19)\\
60344.50 & c & 37.7 (6.2) & 19.96 (0.18)\\
60347.33 & o & 64.1 (7.7) & 19.38 (0.13)\\
60348.55 & c & 40.9 (5.5) & 19.87 (0.14)\\
60349.04 & o & 66.4 (6.7) & 19.34 (0.11)\\
60350.48 & o & 61.4 (6.0) & 19.43 (0.10)\\
60351.18 & o & 66.5 (7.1) & 19.34 (0.11)\\
60353.06 & o & 55.0 (6.1) & 19.54 (0.12)\\
60355.26 & o & 50.6 (6.3) & 19.63 (0.13)\\
60356.95 & o & 55.6 (8.1) & 19.53 (0.16)\\
60361.03 & o & 47.4 (6.4) & 19.71 (0.14)\\
		\hline
\end{longtable}
\end{center}

\begin{center}
\begin{longtable}{lcccccc}
\caption{Log of SN 2022ywf $BVgri$ photometry.}\\
	\label{tab:22ywf_phot}
\textbf{MJD} & \textbf{B} & \textbf{V} & \textbf{g} & \textbf{r} & \textbf{i} & \textbf{Observatory} \\ \hline 
		\hline
59880.84	&	19.05 (0.05)	&	19.80 (0.16)	&	19.15 (0.04)	&	19.50 (0.10)	&	19.51 (0.14)	& LCO \\
59881.86	&	18.61 (0.04)	&	18.91 (0.07)	&	18.70 (0.03)	&	18.77 (0.06)	&	19.02 (0.10)	& LCO \\
59885.15	&	18.21 (0.03)	&	18.32 (0.03)	&	18.17 (0.01)	&	18.28 (0.02)	&	18.50 (0.03)	& LCO \\
59886.25	&	18.17 (0.03)	&	18.30 (0.03)	&	18.14 (0.02)	&	18.24 (0.02)	&	18.50 (0.04)	& LCO \\
59887.99	&	18.40 (0.05)	&	18.32 (0.05)	&	18.33 (0.09)	&	18.25 (0.05)	&	18.43 (0.06)	& LCO \\
59891.80	&	- (-)	&	- (-)	&	- (-)	&	18.14 (0.16)	&	18.10 (0.11)	& LCO \\
59892.50	&	- (-)	&	18.47 (0.07)	&	18.63 (0.05)	&	18.15 (0.06)	&	18.15 (0.07)	& LCO \\
59893.80	&	18.95 (0.09)	&	18.62 (0.12)	&	18.80 (0.06)	&	18.14 (0.05)	&	18.28 (0.08)	& LCO \\
59900.12	&	19.80 (0.05)	&	19.35 (0.05)	&	19.60 (0.04)	&	18.44 (0.03)	&	18.51 (0.05)	& LCO \\
59903.18	&	20.11 (0.06)	&	19.51 (0.06)	&	19.91 (0.04)	&	18.80 (0.03)	&	18.70 (0.04)	& LCO \\
59906.17	&	20.03 (0.06)	&	19.49 (0.05)	&	20.09 (0.04)	&	19.05 (0.03)	&	18.69 (0.05)	& LCO \\
59908.49	&	20.43 (0.08)	&	19.66 (0.06)	&	20.12 (0.04)	&	19.16 (0.03)	&	19.01 (0.04)	& LCO \\
59919.12	&	20.74 (0.38)	&	20.11 (0.33)	&	20.40 (0.30)	&	19.38 (0.14)	&	19.83 (0.27)	& LCO \\
59923.13	&	20.35 (0.15)   &    20.02 (0.16)   &   20.55 (0.18)   &	19.71 (0.11)	&	19.88(0.16)	& LCO \\
59927.05	&	20.18 (0.08)   &    - (-)   &   20.29 (0.07)   & 	- (-)	&	19.67 (0.09)	& LCO \\
59931.07	&	20.69 (0.08)   &    20.97 (0.17)   &   20.93 (0.09)   &	20.05 (0.05)	&	20.05 (0.12)	& LCO \\
59936.90	&	21.02 (0.12)   &    20.93 (0.19)   &   20.79 (0.09)   &	20.37 (0.11)	&	20.03 (0.09)	& LCO \\
59937.95	&	- (-)   &    - (-)   &   20.97 (0.18)   &	20.07 (0.12)	&	20.27 (0.25)	& LCO \\
59960.45	&	- (-)	&    - (-)   &	 21.32 (0.17)  &	20.22 (0.08)	&	19.97 (0.13)	& LCO \\
59973.12	&	- (-)   &    - (-)   &	 - (-)   &	21.02 (0.54)	&	20.14 (0.23)	& LCO \\
59984.09	&	- (-)   &	 - (-)   &	 21.51 (0.17)   &	20.65 (0.13)	&	20.34 (0.14)	& LCO \\
		\hline
\end{longtable}
\end{center}

\newpage
\begin{center}
\begin{longtable}{cccccc} 
\caption{Log of the spectra of SN 2022ywf. } \label{tab:22ywf_spectra} \\
\hline
\multicolumn{1}{c}{\textbf{Date}} & \multicolumn{1}{c}{\textbf{MJD}} & \multicolumn{1}{c}{\textbf{t$_\mathrm{exp}$ [d]}} & \multicolumn{1}{c}{\textbf{Phase [d]}} & \multicolumn{1}{c}{\textbf{Telescope/Instrument}} & \multicolumn{1}{c}{\textbf{Wavelength range [\r{A}]}} \\ \hline 
        2022-10-29 & 59881.3 & 2.9 & -9.1 & APO/Kosmos & 3500 - 6500\\
        2022-10-29 & 59881.5 & 3.1 & -8.9 & LCO/FLOYDS & 3500 - 10000\\
	2022-10-31 & 59883.8 & 5.4 & -6.6 & SALT/RSS & 3500 - 7250\\
        2022-11-01 & 59884.1 & 5.7 & -6.3 & SOAR/GHTS\_RED & 3700 - 7050\\
	2022-11-02 & 59885.9 & 7.5 & -4.5 & SALT/RSS & 3500 - 7250\\
	2022-11-04 & 59887.1 & 8.7 & -3.3 & SOAR/GHTS\_RED & 5000 - 9000\\
	2022-11-05 & 59888.2 & 9.8 & -2.2 & HET/LRS & 3650 - 10000\\
	2022-11-08 & 59891.8 & 13.4 & +1.4 & SALT/RSS & 3500 - 7250\\
	2022-11-09 & 59892.8 & 14.4 & +2.4 & SALT/RSS & 3500 - 7250\\
	2025-11-10 & 59893.2 & 14.8 & +2.8 & HET/LRS & 3700 - 10000\\
    2025-11-10 & 59894.0 & 15.6 & +3.6 & LT/SPRAT & 4050 - 8000\\
    2025-11-12 & 59895.2 & 16.8 & +4.8 & HET/LRS & 3700 - 6900\\
	2025-11-13 & 59896.3 & 17.9 & +5.9 & HET/LRS & 6400 - 10000\\
    2022-11-13 & 59896.4 & 18.0 & +6.0 & LCO/FLOYDS & 3500 - 10000\\
    2025-11-13 & 59897.0 & 18.6 & +6.6 & LT/SPRAT & 4050 - 8000\\
        2022-11-16 & 59899.2 & 20.8 & +8.8 & Bok/B\&C & 4000 - 8000\\
        2022-11-16 & 59899.3 & 20.9 & +8.9 & NTT/EFOSC2 & 3800 - 9300\\
        2022-11-16 & 59899.4 & 21.0 & +9.0 & LCO/FLOYDS & 3500 - 10000\\
		\hline
		\hline
\end{longtable}
\tablefoot{$t_\mathrm{exp}$ shows the time since the date of explosion constrained in the abundance tomography (MJD 59878.4); while the phases are given relative to the maximum in r-band (MJD 59890.4).}
\end{center}

\begin{center}
\begin{longtable}{lccccccr}
\caption{Log of LCO photometry of SN 2023zgx}\\
	\label{tab:23zgx_phot}
\textbf{MJD} & \textbf{B} & \textbf{V} & \textbf{g} & \textbf{r} & \textbf{i} & \textbf{z} &  \textbf{Observatory} \\ \hline
		\hline
60294.12 & 18.95 (0.26) & 17.90 (0.10) & 18.11 (0.11) & 17.48 (0.09) & 17.47 (0.09) & 17.33 (0.15) & RC80 \\ 
60294.15 & 18.75 (0.20) & 17.70 (0.12) & 18.16 (0.12) & 17.40 (0.12) & 17.35 (0.10) & 17.07 (0.16) & BRC80 \\ 
60294.73 & 18.95 (0.06) & 17.81 (0.03) & - (-) & - (-) & - (-) & - (-) & LCO \\ 
60296.13 & 19.07 (0.31) & 18.14 (0.13) & 18.33 (0.16) & 17.71 (0.10) & 17.61 (0.12) & 17.37 (0.24) & RC80 \\ 
60296.17 & 19.00 (0.25) & 17.69 (0.18) & 18.20 (0.16) & 17.58 (0.12) & 17.51 (0.10) & 17.42 (0.19) & BRC80 \\ 
60296.35 & 19.13 (0.07) & 17.94 (0.03) & 18.42 (0.02) & 17.66 (0.02) & 17.61 (0.02) & - (-) & LCO \\ 
60297.11 & 19.37 (0.29) & 18.20 (0.11) & 18.54 (0.14) & 17.85 (0.11) & 17.77 (0.12) & 17.52 (0.14) & RC80 \\ 
60298.13 & 19.49 (0.27) & 18.13 (0.10) & 18.36 (0.12) & 17.83 (0.08) & 17.75 (0.09) & 17.42 (0.15) & RC80 \\ 
60298.33 & 19.18 (0.07) & 18.06 (0.04) & 18.50 (0.03) & 17.81 (0.02) & 17.69 (0.03) & - (-) & LCO \\ 
60300.09 & 19.03 (0.37) & 18.02 (0.09) & 18.77 (0.10) & 17.90 (0.07) & - (-) & - (-) & LCO \\ 
60304.35 & 19.12 (0.36) & 18.33 (0.29) & 18.80 (0.10) & 18.17 (0.06) & 17.88 (0.11) & - (-) & LCO \\ 
60305.44 & 19.86 (0.60) & 18.68 (0.19) & 18.64 (0.12) & 18.20 (0.09) & 18.21 (0.08) & - (-) & LCO \\ 
60307.07 & 19.55 (0.23) & 18.52 (0.11) & 19.04 (0.10) & 18.37 (0.06) & 18.29 (0.07) & - (-) & BRC80 \\ 
60308.20 & 20.04 (0.28) & 18.81 (0.12) & 19.16 (0.08) & 18.36 (0.06) & 18.35 (0.07) & - (-) & LCO \\ 
60309.07 & - (-) & 18.87 (0.30) & 19.08 (0.27) & 18.56 (0.12) & 18.36 (0.12) & 17.89 (0.17) & RC80 \\ 
60309.25 & 19.83 (0.17) & 18.54 (0.08) & 18.81 (0.05) & 18.33 (0.05) & 18.20 (0.06) & - (-) & LCO \\ 
60310.43 & 19.80 (0.17) & 18.74 (0.09) & 19.48 (0.13) & 18.42 (0.06) & 18.32 (0.08) & - (-) & LCO \\ 
60315.51 & 20.41 (0.18) & 18.93 (0.05) & 19.33 (0.05) & 18.76 (0.04) & 18.52 (0.04) & - (-) & LCO \\ 
60319.04 & - (-) & 19.20 (0.19) & 19.52 (0.17) & 18.78 (0.14) & 18.70 (0.19) & - (-) & RC80 \\ 
60320.05 & - (-) & 19.13 (0.16) & 19.76 (0.21) & 19.02 (0.12) & - (-) & - (-) & RC80 \\ 
60322.05 & - (-) & - (-) & - (-) & 18.94 (0.21) & 18.62 (0.22) & - (-) & RC80 \\ 
60322.68 & 20.40 (0.12) & 19.28 (0.06) & 19.64 (0.07) & 19.42 (0.25) & 18.60 (0.05) & - (-) & LCO \\ 
60326.04 & - (-) & 19.24 (0.23) & 19.81 (0.28) & - (-) & 18.65 (0.18) & - (-) & RC80 \\ 
60328.35 & 20.52 (0.06) & 19.35 (0.04) & 19.80 (0.04) & 19.13 (0.03) & 18.79 (0.03) & - (-) & LCO \\ 
60329.04 & - (-) & 19.04 (0.37) & - (-) & 19.00 (0.20) & - (-) & 18.20 (0.34) & RC80 \\ 
60330.02 & - (-) & - (-) & - (-) & 19.13 (0.26) & 19.05 (0.28) & - (-) & RC80 \\ 
60331.09 & - (-) & 19.16 (0.33) & 19.41 (0.37) & 19.13 (0.20) & 18.69 (0.17) & 18.82 (0.36) & RC80 \\ 
60332.32 & 20.59 (0.13) & 19.45 (0.06) & 19.93 (0.06) & 19.27 (0.05) & 18.84 (0.03) & - (-) & LCO \\ 
60333.08 & - (-) & - (-) & 19.45 (0.38) & 19.13 (0.17) & - (-) & - (-) & RC80 \\ 
60335.02 & - (-) & 19.37 (0.25) & - (-) & - (-) & 18.75 (0.26) & - (-) & RC80 \\ 
60336.15 & 20.71 (0.42) & 19.38 (0.16) & 20.10 (0.21) & 19.28 (0.12) & 19.05 (0.11) & - (-) & LCO \\ 
60338.09 & - (-) & - (-) & 20.38 (0.61) & 19.27 (0.16) & 18.62 (0.14) & - (-) & BRC80 \\
60342.29 & 21.07 (0.26) & 19.76 (0.11) & 20.10 (0.11) & 19.32 (0.06) & 18.96 (0.06) & - (-) & LCO \\ 
60346.11 & 20.40 (0.21) & 19.72 (0.12) & 19.81 (0.09) & 18.99 (0.05) & 19.06 (0.08) & 18.93 (0.20) & BRC80 \\
60347.22 & 20.92 (0.11) & 19.76 (0.06) & 20.27 (0.06) & 19.52 (0.04) & 19.09 (0.04) & - (-) & LCO \\ 
60350.93 & 21.03 (0.23) & 20.19 (0.14) & 20.33 (0.08) & 19.70 (0.07) & 19.07 (0.08) & - (-) & LCO \\ 
60355.09 & 21.38 (0.32) & 19.96 (0.11) & 20.58 (0.09) & 19.77 (0.08) & 19.54 (0.10) & - (-) & LCO \\ 
60357.02 & - (-) & 20.01 (0.19) & 20.23 (0.15) & 19.69 (0.11) & 19.30 (0.10) & 18.52 - (-) & BRC80 \\
60359.11 & 21.51 (0.24) & 20.39 (0.13) & 20.66 (0.10) & 19.98 (0.08) & 19.26 (0.08) & - (-) & LCO \\ 
60362.94 & 20.61 (0.41) & 20.00 (0.24) & 20.01 (0.20) & 20.21 (0.37) & 19.35 (0.18) & - (-) & LCO \\ 
60369.97 & 21.34 (0.64) & 20.34 (0.33) & 20.38 (0.26) & 19.95 (0.21) & 19.27 (0.12) & - (-) & LCO \\ 
60374.26 & 21.14 (0.24) & 21.11 (0.29) & 20.83 (0.13) & 20.22 (0.08) & 19.51 (0.06) & - (-) & LCO \\ 
60378.18 & 21.94 (0.27) & 20.52 (0.12) & 20.95 (0.12) & 20.23 (0.08) & 19.54 (0.06) & - (-) & LCO \\ 
60382.13 & 21.79 (0.31) & 20.80 (0.20) & 21.44 (0.25) & 20.43 (0.15) & 19.67 (0.13) & - (-) & LCO \\ 
60387.84 & 22.22 (0.91) & 20.77 (0.45) & 21.61 (0.48) & 20.56 (0.14) & 19.71 (0.08) & - (-) & LCO \\ 
60389.03 & - (-) & 20.55 (0.48) & 20.93 (0.47) & 19.88 (0.17) & 19.56 (0.17) & - (-) & BRC80 \\
60392.09 & - (-) & 20.48 (1.09) & 19.75 (0.60) & 20.36 (0.22) & 19.10 (0.15) & - (-) & LCO \\ 
60397.25 & 23.16) (2.07) & 20.98 (0.45) & - (-) & - (-) & - (-) & - (-) & LCO \\ 
60404.90 & 22.48 (0.52) & 21.05 (0.21) & 21.93 (0.32) & 21.05 (0.25) & 20.00 (0.16) & - (-) & LCO \\ 
60416.92 & 23.76 (2.19) & 21.84 (0.56) & 22.70 (0.79) & 21.75 (0.48) & 20.88 (0.29) & - (-) & LCO \\ 
60424.80 & - (-) & - (-) & 21.60 (1.65) & 20.95 (0.82) & 20.21 (0.43) & - (-) & LCO \\ 
60432.97 & 23.50) (1.02) & 22.76 (0.80) & - (-) & - (-) & - (-) & - (-) & LCO \\ 
60440.98 & - (-)) & 22.25 (0.65) & - (-) & - (-) & - (-) & - (-) & LCO \\ 
60451.80 & - (-) & - (-) & 21.22) (1.07) & 20.56 (0.50) & 20.12 (0.36) & - (-) & LCO \\ 
		\hline
\end{longtable}
\end{center}

\begin{center}
\begin{longtable}{cccccc} 
\caption{Log of the spectra of SN 2023zgx } \label{tab:23zgx_spectra} \\
\hline
\multicolumn{1}{c}{\textbf{Date}} & \multicolumn{1}{c}{\textbf{MJD}} & \multicolumn{1}{c}{\textbf{t$_\mathrm{exp}$ [d]}} & \multicolumn{1}{c}{\textbf{Phase [d]}} & \multicolumn{1}{c}{\textbf{Telescope/Instrument}} & \multicolumn{1}{c}{\textbf{Wavelength range [\r{A}]}} \\ \hline 
		2023-12-13 & 60291.6 & 15.5 & +1.2 & Gemini-N/GMOS & 4700 - 9250\\
		2023-12-16 & 60294.3 & 18.2 & +3.9 & NTT/EFOSC2 & 3650 - 9220\\
		2023-12-19 & 60297.3 & 21.2 & +6.9 & NTT/EFOSC2 & 3650 - 9220\\
		2023-12-31 & 60309.3 & 33.2 & +18.9 & NTT/EFOSC2 & 3650 - 9220\\
            2024-01-06 & 60315.6 & 39.5 & +25.2 & LCO/FLOYDS & 3500-10000\\
            2024-01-13 & 60322.7 & 46.6 & +32.3 & LCO/FLOYDS & 3500-10000\\
            2024-01-16 & 60325.3 & 49.2 & +34.9 & NTT/EFOSC2 & 3650 - 9220\\
            2024-02-04 & 60344.6 & 68.5 & +54.2 & LCO/FLOYDS & 3500 - 10000\\
            2024-02-12 & 60352.5 & 76.4 & +62.1 & LCO/FLOYDS & 3500 - 10000\\
            2024-02-20 & 60360.6 & 84.5 & +70.2 & LCO/FLOYDS & 3500 - 10000\\
            2024-03-01 & 60370.9 & 94.8 & +80.5 & SALT/RSS & 3500 - 7250\\
            2024-03-02 & 60371.9 & 95.8 & +81.5 & SALT/RSS & 3500 - 7250\\
            2024-03-10 & 60379.0 & 103.9 & +89.6 & SALT/RSS & 3500 - 7250\\
            2024-03-11 & 60380.0 & 104.9 & +90.6 & SALT/RSS & 3500 - 7250\\
		\hline
		\hline
\end{longtable}
\tablefoot{$t_\mathrm{exp}$ shows the time since the date of explosion constrained in the abundance tomography (MJD 60276.1); while the phases are given relative to the maximum in r-band (MJD 60290.4).}
\end{center}

\begin{center}
\begin{longtable}{c | cc | cc} 
\caption{The peak absolute magnitudes in r/R-band and photospheric velocities of type Iax SNe at the moment of maximum light.} \label{tab:citations} \\
\hline
\multicolumn{1}{c}{\textbf{Supernova}} & \multicolumn{2}{l}{\textbf{  $M_\mathrm{r}$}} & \multicolumn{2}{l}{\textbf{  $v_\mathrm{phot}$}}  \\ \hline 
		SN 2011ay & -18.60 &  \citet{Szalai15} & 8100 & \citet{Barna17}\\
        SN 2012Z  & -18.60 &  \citet{Stritzinger15} & 7600 & \citet{Barna18}\\
        SN 2005hk & -18.07 &  \citet{Stritzinger15} & 6700 & \citet{Barna18}\\
        SN 2020udy & -17.86 &  \citet{Maguire23} & 7900 & \citet{Singh24}\\
        SN 2018cni & -17.82 &  \citet{Singh23} & 7500 & \citet{Singh23}\\
        SN 2002cx & -17.67 &  \citet{Li03} & 6700 & \citet{Barna18}\\
        SN 2015H & -17.27 &  \citet{Magee16} & 6100 & \citet{Barna18}\\
        SN 2019muj & -16.40 &  \citet{Barna21} & 4600 & \citet{Barna21}\\
        SN 2022xlp & -16.06 &  \citet{Banhidi25} & 4700 & \citet{Banhidi25}\\          
        SN 2023zgx & -14.70 &  This work & 3100 & This work \\
        SN 2020kyg & -14.50 &  \citet{Singh23} & 3000 & \citet{Singh23}\\ 
        SN 2019gsc & -13.97 &  \citet{Srivastav20} & 2100 & \citet{Srivastav20} \\
        SN 2022ywf & -13.81 &  This work & 2400 & This work\\ 
            \hline
		\hline
\end{longtable}
\end{center}


\label{lastpage}
\end{document}